\documentclass[aps,prd,twocolumn,showpacs,superscriptaddress,nofootinbib,preprintnumbers]{revtex4-2}

\usepackage{graphicx}
\usepackage{color}
\usepackage{amsmath, bm}
\usepackage{amssymb}
\usepackage{enumerate}
\usepackage[colorlinks=true,citecolor=blue,urlcolor=blue]{hyperref}
\usepackage[capitalise]{cleveref}
\usepackage{multirow}
\graphicspath{ {Figures/} }
\begin{document}\
\preprint{UTWI-12-2026}

\title{CMB constraints on dark matter-proton scattering: investigating prior-volume effects using profile likelihoods}

\begin{abstract}
We present profile-likelihood constraints on velocity-independent dark matter-proton scattering, including cases in which only a fraction of dark matter has such non-gravitational interactions. 
Frequentist profile-likelihood techniques provide prior-independent constraints, circumventing prior-volume effects that we show arise in Bayesian constraints on this model.
In the limit where the scattering cross section or the fraction of interacting dark matter approaches zero, 
the other interacting dark matter model parameters become unconstrained, causing the posterior distribution to favor that region of parameter space. 
Using \textit{Planck} 2018 cosmic microwave background anisotropy data, we find a clear impact of prior-volume effects on the posteriors used to place constraints on dark matter scattering. 
Compared to the frequentist analysis, the Bayesian method consistently overestimates the constraints on the cross section. 
Given the potentially biased upper limits on models subject to prior-volume effects, such as this one, we recommend supplementing Bayesian constraints with frequentist statistics to better assess the impact of priors.
\end{abstract}

\author{Maria C. Straight}\thanks{NSF Graduate Research Fellow}\thanks{maria.straight@utexas.edu}
\affiliation{Department of Astronomy, The University of Texas at Austin, Austin, TX 78712, USA}
\affiliation{Texas Center for Cosmology and Astroparticle Physics, Weinberg Institute, Department of Physics, The University of Texas at Austin, Austin, TX 78712, USA}

\author{Tanvi Karwal}
\affiliation{Kavli Institute for Cosmological Physics, Enrico Fermi Institute, and Department of Astronomy \& Astrophysics, University of Chicago, Chicago, IL 60637, USA}

\author{Jos\'{e} Luis Bernal}
\affiliation{Instituto de F{\'i}sica de Cantabria (IFCA), CSIC-Univ. de Cantabria, Avda. de los Castros s/n, Santander, E-39005, Spain}

\author{Kimberly K.~Boddy}
\affiliation{Texas Center for Cosmology and Astroparticle Physics, Weinberg Institute, Department of Physics, The University of Texas at Austin, Austin, TX 78712, USA}

\maketitle

%%%%%%%%%%%%%%%%%%%%%%%%%%%%%%%%%%%%%%%%%%%%%%%%%%%%%%%%%%%%%%%%%%%%%%%%%%%%%%%

\section{Introduction}

Within the standard $\Lambda$-Cold Dark Matter ($\Lambda$CDM) cosmological model, dark matter is modeled as a cold, collisionless fluid that only interacts gravitationally. While $\Lambda$CDM succeeds in describing several large-scale observations of the Universe \citep{PlanckCollaboration_2020_Planck,Miyatake:2023njf,SPT_2024,AtacamaCosmologyTelescope_2025,Wright:2025xka,DES_2026,DES:2026mkc}, it only provides a heuristic description of dark matter. 
In an effort to understand the microscopic properties of dark matter, there are dedicated efforts to search for non-gravitational interactions between dark matter and the known fundamental particles within the Standard Model.

Cosmological observations provide probes of non-gravitational interactions of dark matter that are complementary to particle physics searches, such as direct detection experiments \citep{Cushman_2013_snowmass,XENON_2023_First,PandaX_2024_Dark,LZ_2024_Dark}. Elastic scattering between dark matter and baryons in the dense early universe reduces the amplitude of dark matter density fluctuations and suppresses power on small scales in the cosmic microwave background (CMB) power spectrum. Precise measurements of the CMB can probe a wider range of dark matter masses than direct detection experiments \citep{Cushman_2013_snowmass} and have no upper limit on the cross sections they can constrain \citep{Emken_2017_DaMaSCUS,BoddyGluscevic_2018_first}.

Limits on dark matter interactions with baryons from CMB anisotropies have previously been obtained for velocity-independent and velocity-dependent scattering
\citep{Chen_2002_cosmic, Dvorkin_2013_constraining,Boddy_2018_new,GluscevicBoddy_2018_constraints,Xu_2018_Probing, Nguyen_2021_observational, Buen-Abad_2022_Cosmological,Boddy_2022_Investigation,Li:2022mdj, An_2024_Interacting}.
Suppression of structure formation due to dark matter scattering with baryons has also been studied in the context of the Lyman-$\alpha$ forest \citep{Dvorkin_2013_constraining,Munoz_2017_Constraints,Xu_2018_Probing,Ooba_2019_Cosmological,Rogers_2021_Limits}, 
the 21-cm signal \citep{Munoz_2015_heating,Slatyer_2018_Early,Rahimieh_2025_sensitivity,Rahimieh_2025_Forecasting,Kovetz_2018_Tighter,Short_2022_Dark,Driskell_2022_Structure}, 
weak lensing \citep{Zhang_2024_weak,He_2025_bounds}, 
the Milky-Way satellite abundance \citep{Nadler_2019_Constraints,Maamari_2020_Bounds,Nguyen_2021_observational,Nadler_2025_COZMIC}, 
and the galaxy ultraviolet luminosity function \citep{Lazare_2025_First,Das:2025bnr}. 
However, fewer analyses have considered fractional cases in which only some of the dark matter interacts with baryons \citep{Boddy_2018_new,He_2023_S8,He_2025_bounds}. Additionally, most of these analyses employed Bayesian inference using numerical techniques such as Markov chain Monte Carlo (MCMC) sampling.

The parameterization capturing physics beyond cold dark matter is often prone to prior-volume effects in Bayesian analyses. 
For example, the cosmological effect of dark matter-baryon scattering depends on two main parameters: the momentum-transfer cross section and the dark matter particle mass. As the cross section goes to zero, the mass can take any possible value within the chosen prior range without having any impact on the predicted observables. Additionally, considering a scenario in which only a fraction of the total dark matter abundance interacts (while the rest behaves as standard cold dark matter) results in yet another parameter contributing to the prior-volume effects. As the fraction goes to zero, the mass and the cross section can both take any value within their prior ranges without affecting the likelihood, creating an even larger acceptable prior volume for a Bayesian MCMC to explore unnecessarily.

According to Bayes's theorem, the posterior is proportional to the product of the likelihood and the prior; thus, the choice of prior may shift the posterior. 
However, we stress the fact that in the case of interacting dark matter, which has no well-motivated priors, the prior's impact on the posterior 
is an artifact rather than a conscious, intentional choice. 
As our prior includes the $\Lambda$CDM limit, the posterior may be artificially shifted towards $\Lambda$CDM where the likelihood flattens and the new model parameters become unconstrained. 
The posterior in this region of the parameter space then simply reflects the chosen prior, potentially resulting in overly strong constraints on new physics. 
Similar prior-volume effects occur in models within and beyond $\Lambda$CDM and have been studied for early dark energy \cite{Smith_2020_Early,Herold_2021_New,Herold:2022iib,Karwal_2024_Procoli}, effective field theories of large-scale structure \cite{Holm:2023laa, Zhang:2024thl, Chudaykin:2024wlw}, and intrinsic alignment nuisance parameters in cosmic shear analyses \cite{DES:2021vln,Kilo-DegreeSurvey:2023gfr,Arico:2023ocu}. 

The dark matter-baryon scattering parameter space has been explored in previous work by fixing one or two of the three new dark matter-baryon scattering parameters, while allowing other dark matter and $\Lambda$CDM parameters to vary in a Bayesian analysis. 
However, the prior-volume effects and any subsequent biases can be especially impactful when all dark matter parameters are varied simultaneously. 
Bayes factor surfaces \citep{Fowlie_2024_Bayes} or frequentist methods---such as profile likelihoods \citep{Herold_2024_profile}---avoid this problem, since they do not rely on using prior distributions. 
Profile-likelihood constraints on dark matter-proton scattering have been obtained using weak-lensing data, under the assumption that 100\% of the dark matter scatters with protons \citep{Zhang_2024_weak}.

In this work, we present CMB profile-likelihood constraints on velocity-independent dark matter-proton scattering, as well as the first profile-likelihood constraints on fractional dark matter-proton scattering. Our work is also the first to include an analysis of this model in which the dark matter mass, scattering cross section, and interaction fraction are simultaneously varied within an MCMC. We perform a detailed comparison between frequentist profile-likelihood and Bayesian MCMC constraints, and we explore how different priors for the cross section and dark matter interaction fraction impact the constraints due to prior-volume effects. Compared to a profile-likelihood analysis, the MCMC constraints on the cross section for fixed interaction fraction are more stringent, with the most stringent constraints coming from analyses with the broadest priors that extend the farthest into the $\Lambda$CDM-consistent regions of parameter space.

The paper is organized as follows. In Sec.~\ref{sec:scattering}, we introduce the modified Boltzmann equations that incorporate the physics of dark matter-proton scattering. In Sec.~\ref{sec:Bayesian&ProfLkl}, we describe the Bayesian and frequentist statistical frameworks that we employ for our study. In Sec.~\ref{sec:Methods}, we describe the data and codes we use for our analysis. We present our constraints, compare the MCMC and profile-likelihood results, and explore the impact of prior-volume effects in Sec.~\ref{sec:Results}. We conclude in Sec.~\ref{sec:Conclusions}.

\section{Dark matter-proton scattering}\label{sec:scattering}

Elastic scattering between dark matter and protons in the early universe introduces heat and momentum exchange between the dark matter and baryon fluids. The relevant quantity appearing in the linear Boltzmann equations that captures the effect of scattering is the momentum-transfer cross section
\begin{equation}
    \sigma_\chi = \int{\mathrm{d}\Omega \frac{\mathrm{d}\sigma}{\mathrm{d}\Omega} (1-\cos{\theta})} \,,
    \label{eq:sigma}
\end{equation}
where $\mathrm{d}\sigma/\mathrm{d}\Omega$ is the differential cross section, and the integral over solid angle is weighted by the fractional longitudinal momentum $1-\cos\theta$.
This weighting suppresses far-forward scattering, giving preference to larger scattering angles $\theta$, which induce larger transfers of momentum between the scattering particles.
The momentum-transfer cross section can be parameterized as $\sigma_\chi = \sigma_0 v^n$ for a broad class of dark matter models \citep{BoddyGluscevic_2018_first}, where $v$ is the relative particle velocity between protons and dark matter.

In this work, we focus on the velocity-independent case in which $n=0$, and the momentum-transfer cross section $\sigma_0$ coincides with the standard cross section [i.e., $\int d\Omega (d \sigma/ d\Omega)$].
Two common velocity-independent scattering models correspond to spin-independent and spin-dependent interactions, the standards for reporting direct detection results.
Dark matter scattering with protons (either free or within neutral hydrogen atoms) in the early universe can also imply scattering with the protons in helium nuclei, unless the interaction is spin-dependent~\citep{GluscevicBoddy_2018_constraints,BoddyGluscevic_2018_first}.
For spin-independent interactions, incorporating helium scattering would improve constraining power, particularly for dark matter masses above a GeV \citep{Boddy_2022_Investigation}. However, to avoid possible model-dependent caveats, we consider the more conservative case of dark matter scattering with neutral/ionized hydrogen only. We expect the main conclusions of this work to apply similarly to variants of the model considered.

The scattering process introduces a collision term in the Boltzmann equations for the evolution of the dark matter and baryon fluctuations.
In synchronous gauge \citep{Ma_1995_cosmological}, the dark matter and baryon density fluctuations $\delta_\chi$ and $\delta_b$, respectively, and corresponding velocity divergences $\theta_\chi$ and $\theta_b$ evolve as
\begin{subequations}
\begin{equation}
    \dot{\delta}_\chi = -\theta_\chi - \frac{\dot{h}}{2} \,,
\end{equation} 
\begin{equation}
    \dot{\delta}_b = -\theta_b - \frac{\dot{h}}{2} \,,
\end{equation} 
\begin{equation}
    \dot{\theta}_\chi = -\frac{\dot{a}}{a}\theta_\chi + c_\chi^2 k^2 \delta_\chi + R_\chi(\theta_b - \theta_\chi) \,, \,\, \mathrm{ and }
\end{equation}
\begin{equation}
    \dot{\theta}_b = -\frac{\dot{a}}{a}\theta_b + c_b^2 k^2 \delta_b + R_\gamma(\theta_\gamma - \theta_b) + \frac{\rho_\chi}{\rho_b} R_\chi(\theta_\chi-\theta_b) \,,
\label{eq:Boltzmann_d}
\end{equation}
\label{eq:Boltzmann}%
\end{subequations}
where $c_\chi$ and $c_b$ are the adiabatic sound speeds, $\rho_\chi$ and $\rho_b$ are the background energy densities, $k$ is the Fourier-mode wave number, $a$ is the scale factor, $h$ is the trace of the scalar metric perturbation, and overdots denote a derivative with respect to conformal time. $R_\gamma=(4/3)(\rho_\gamma/\rho_b)an_e\sigma_T$ is the coefficient of the momentum-transfer rate between baryons and photons from Thomson scattering \citep{Ma_1995_cosmological}, and $R_\chi$ is the coefficient for the rate of momentum exchange between dark matter and baryons.

The dark matter-baryon momentum-exchange rate coefficient is
\begin{equation}
    R_\chi = \mathcal{N}_0 a \rho_b (1-Y_\mathrm{He})\frac{\sigma_\chi}{m_\chi+m_p} \Bigg( \frac{T_b}{m_p}+\frac{T_\chi}{m_\chi}\Bigg)^{\frac{1}{2}} \,,
    \label{eq:Rchi}
\end{equation}
where $\mathcal{N}_0\equiv 2^{7/2}/ (3\sqrt{\pi})$, $Y_\mathrm{He}$ is the helium mass fraction, $m_p$ and $m_\chi$ are the proton and dark matter masses, 
and $T_b$ and $T_\chi$ are the temperatures of the baryon and dark matter fluids, respectively. 
These temperatures evolve as
\begin{subequations}
\begin{equation}
    \dot{T}_\chi = -2\frac{\dot{a}}{a}T_\chi + 2R'_\chi(T_b-T_\chi) \,,
\label{eq:temp_evol_a}
\end{equation}
\begin{equation}
    \dot{T}_b = -2\frac{\dot{a}}{a}T_b + \frac{2\mu_b}{m_\chi}\frac{\rho_\chi}{\rho_b}R'_\chi(T_\chi-T_b) + \frac{2\mu_b}{m_e}R_\gamma(T_\gamma-T_b) \,,
\label{eq:temp_evol_b}
\end{equation} \label{eq:temp_evol}%
\end{subequations}
where $m_e$ is the electron mass; $\mu_b\approx m_p(n_\mathrm{H}+4n_\mathrm{He})/(n_\mathrm{H}+n_\mathrm{He}+n_e)$ is the mean molecular weight of the baryons; $n_\mathrm{H}$, $n_\mathrm{He}$, and $n_\mathrm{e}$ are the number densities of hydrogen, helium, and electrons; and the heat-exchange rate coefficient is $R'_\chi=R_\chi m_\chi/(m_\chi+m_p)$.
We note that the peculiar velocity between dark matter and baryons does generally impact the evolution equations, but the effect is completely negligible for $n\geq 0$ at the level of CMB constraints~\citep{Dvorkin_2013_constraining,Boddy_2018_new,Ali-Haimoud_2024_exact}.

These evolution equations are further modified when only a fraction $f_\chi$ of the total dark matter abundance interacts with baryons. To account for the fractional case, we replace $\rho_\chi \rightarrow f_\chi \rho_\chi$ in Eq.~(\ref{eq:Boltzmann_d}) and Eq.~(\ref{eq:temp_evol_b}).
We assume the remaining $1-f_\chi$ is standard cold dark matter.

\section{Bayesian posteriors and frequentist profile likelihoods}\label{sec:Bayesian&ProfLkl}
Bayesian statistics is commonly used for parameter inference in cosmology, typically by employing an MCMC sampler to obtain the posterior. The posterior $\mathcal{P}(\bm\theta|\bm d)$ describes the probability of the parameters $\bm \theta$ given the data $\bm d$ according to Bayes' theorem, 
\begin{equation}
    \mathcal{P}(\bm \theta|\bm d)\propto\Pi(\bm \theta)\mathcal{L}(\bm d|\bm \theta) \,,
    \label{eqn:Bayesian}
\end{equation}
where the likelihood $\mathcal{L}(\bm d|\bm \theta)$ is the probability of the data $\bm d$ given the specific parameter values $\boldsymbol{\theta}$ and 
the prior distribution $\Pi(\bm \theta)$ incorporates prior knowledge about the expected values of the parameters. 
Bayesian credible intervals then express a probability or degree of belief that a parameter value lies within a certain range given the observed data and prior belief.

In cases where there is no robust argument for a strong prior, the posterior distribution generated by an MCMC analysis should be determined by the likelihood, and we can use the marginalized posterior to generate a constraint on a model parameter for the given data set.
However, in the case that the posterior distribution is influenced or driven by a non-motivated prior, the interpretation of the posterior distribution is less clear and may result in biased conclusions.

In this work, we consider how prior-volume effects can influence marginal posterior distributions in the context of dark matter-proton scattering. This model contains two parameters that may introduce prior-volume effects into the analysis: the cross section $\sigma_0$ and the interaction fraction $f_\chi$. 
In the absence of evidence for dark matter scattering, the marginal posterior distribution could set an upper limit on either of these parameters (both have physical lower limits of 0) by finding the value of the parameter at the edge of the 95.45\% credible interval.

However, as either $\sigma_0$ or $f_\chi$ approaches zero, the model becomes indistinguishable from $\Lambda$CDM, and the properties defined by the other new-physics parameters no longer affect the predicted observables. In this region of parameter space where the likelihood is flat, a Bayesian analysis, by design, simply samples the prior volume of the new-physics parameters. 
For example, in the $\sigma_0 \rightarrow 0$ limit, $f_\chi$ and $m_\chi$ become unconstrained by observables and can vary freely. As a result, the posterior distribution for $\sigma_0$ may favor these small values of $\sigma_0$. The same effect can similarly influence the posterior distribution for $f_\chi$. 

In addition to using a Bayesian MCMC, we employ prior-independent frequentist statistics using profile likelihoods. 
A profile likelihood analysis maximizes the likelihood over all other parameters for given values of the parameters of interest. 
Consider the parameter space $\boldsymbol{\theta}=(\bm \mu, \bm \nu)$, where $\bm \mu$ represents new model parameters of interest, and $\bm \nu$ represents all other cosmological and nuisance parameters. 
The profile likelihood over $\bm{\mu}$ is then
\begin{equation}
    \Delta\chi^2(\bm \mu) = -2 \ln\Bigg(\frac{\mathcal{L}(\bm d|\bm \mu,\Tilde{\bm \nu})}{\mathcal{L}(\bm d|\bm{\hat\mu}, \bm{\hat\nu})}\Bigg) \,,
    \label{eqn:profilelikelihood}
\end{equation}
where $\mathcal{L}(\bm d|\bm{\hat\mu}, \bm{\hat\nu})$ denotes the likelihood of the data at the global maximum likelihood estimate (MLE) with best-fit parameters $\bm{\hat\mu}$ and $\bm{\hat\nu}$, and $\mathcal{L}(\bm d|\bm \mu,\bm \Tilde{\bm \nu})$ is the conditional MLE with optimized parameters $ \Tilde{\bm \nu}$ for fixed values of $\bm{\mu}$. 

In contrast to a Bayesian credible interval, a 95.45\% frequentist confidence interval is defined such that, assuming correct coverage, 95.45\% of intervals constructed from repeated experiments cover the true value of the parameter. 
Confidence intervals can be obtained, for example, by using either the Neyman construction or the approximate graphical method. 
The full Neyman construction \citep{Neyman_1937_Outline} involves generating mock data for all relevant values of $\bm{\theta}$ and evaluating the likelihood. 
Although it guarantees correct coverage, the high computational cost typically makes this approach prohibitive.

The graphical profile-likelihood method provides correct coverage for a Gaussian-distributed parameter or in the asymptotic large-sample limit 
when, according to Wilks’ theorem \citep{Wilks_1938_Large-Sample}, the log-likelihood ratio statistic in \cref{eqn:profilelikelihood} follows a $\chi^2$ distribution with $N$ degrees of freedom for an $N$-dimensional profile likelihood. 
If the parameter follows a Gaussian distribution, i.e., the profile likelihood $\Delta\chi^2(\bm{\mu})$ is parabolic, then the graphical method begins by plotting \cref{eqn:profilelikelihood}. The 68.27\%, 95.45\%, and 99.73\% confidence intervals are defined as the range in $\bm{\mu}$ where the curve is less than a $\Delta\chi^2$ cutoff value of 1 (2.30), 4 (6.18), 9 (11.83), respectively, for profile likelihood with one (two) dimensions. 
Notably, profile likelihoods are invariant under parameter redefinitions, such as the choice of using a linear or logarithmic parameterization. Therefore, this graphical method can be applied as long as there exists a reparameterization in which the profile likelihood is parabolic. 

Using mock \textit{Planck}-lite spectra, Ref.~\citep{Herold_2024_profile} verified that the asymptotic limit is achieved for $\Lambda$CDM, validating the use of Wilks' theorem and indicating that the graphical profile-likelihood construction gives correct coverage. However, extensions to $\Lambda$CDM have new parameters to consider. When a parameter is near a physical boundary, its log-likelihood ratio statistic cannot be $\chi^2$-distributed, and Wilks' theorem no longer holds. In this case, the Feldman-Cousins \citep{Feldman_1997_Unified} prescription of confidence intervals and a boundary-corrected graphical construction are more appropriate, which we discuss further in Appendix~\ref{app:Feldman-Cousins}.

\section{Methods}\label{sec:Methods}
Our analysis uses the \textit{Planck} Likelihood Code (PLC) version R3.10 of the \textit{Planck} 2018 temperature, polarization, and lensing power spectra of the CMB \citep{PlanckCollaboration_2020_Planck}. We use the baseline low-$\ell$ likelihood for TT and EE and the baseline lensing likelihood, which uses the SMICA temperature and polarization map-based lensing reconstruction. For the high-$\ell$ likelihoods, we use the \texttt{Plik lite} nuisance-marginalized joint TT, TE, EE likelihood. 
We use a modified version\footnote{\url{https://github.com/kboddy/class_public/tree/dmeff}} of the \texttt{CLASS} \citep{Blas_2011_CLASS} code that includes dark matter scattering with protons \citep{GluscevicBoddy_2018_constraints} and solves the modified Boltzmann equations described in Sec.~\ref{sec:scattering}. 
We use \texttt{Cobaya} \citep{Torrado_2020_Cobaya} to perform our Bayesian analyses using the Metropolis-Hastings MCMC algorithm. 

The cross section and fraction both may have a large dynamical range spanning several orders of magnitude. For the cross section, for example, the range extends from values arbitrarily close to zero to the current known upper limits (e.g., Refs.~\cite{Boddy_2018_new,Nguyen_2021_observational}). We choose to use logarithmic priors (or, equivalently, uniform priors in logspace) to ensure uniform coverage of the dynamical range these two parameters may span. 
For fixed interaction fractions $\log_{10}(f_\chi) =0$, $-0.5$, $-1$, $-1.5$, and $-2$, we allow the cross section, mass, and other cosmological and nuisance parameters to vary. 
For flat priors on $\log_{10}(\sigma_0/\mathrm{cm}^2)$, we adjust the range to accommodate larger cross sections for smaller interaction fractions as shown in \cref{tab:f_sigma_priors}. 
For every case, we use the same prior for the mass. 
We also adopt broad flat priors on the six $\Lambda$CDM cosmological parameters $\{ \ln(10^{10} A_\mathrm{s}),\, n_\mathrm{s},\, 100\theta_*,\, \Omega_\mathrm{b} h^2,\, \Omega_\mathrm{c} h^2,\, \tau_\mathrm{reio} \}$ and use the recommended prior for the \textit{Planck} absolute-calibration nuisance parameter $A_\mathrm{\text{Planck}}$ \citep{PlanckCollaboration_2020_Planck}. We compare the resulting Bayesian posteriors to the frequentist profile likelihoods.

\begin{table}[]
    \centering
    \begin{tabular}{c|c|c}
        \multirow{2}{*}{$\log_{10}f_\chi$} & Prior on & Prior on \\
         & $\log_{10}(\sigma_0/\mathrm{cm}^2)$ & $\log_{10}(m_\chi/\mathrm{GeV})$ \\
        \hline \hline 
        $0$ & \multirow{2}{*}{$[-27, -21]$} & \multirow{5}{*}{$[-5,3]$}\\
        $-0.5$ & \\
        \cline{1-2}
        $-1$ & \multirow{2}{*}{$[-27, -20]$}\\
        $-1.5$ & \\
        \cline{1-2}
        $-2$ & \multirow{1}{*}{$[-27, -19]$}\\
    \end{tabular}
    \caption{Bayesian flat prior bounds on the interaction cross section $\sigma_0$ and dark matter particle mass $m_\chi$ for various fixed interaction fractions $f_\chi$. 
    We vary the $\sigma_0$ prior only to ensure that the upper limits are captured within the parameter range. 
    }
    \label{tab:f_sigma_priors}
\end{table}

To test how the choice of prior affects the posterior for $\sigma_0$, we also obtain Bayesian constraints for different ranges of the flat prior on $\log_{10}(\sigma_0/\mathrm{cm}^2)$, fixing $f_\chi = 1$. The upper bound of the prior is set to $\log_{10}(\sigma_0/\mathrm{cm}^2) = -21$, while we vary the lower bounds to $-27$, $-29$, and $-31$. Decreasing the lower bound pushes further into the $\Lambda$CDM-like region of parameter space and increases the amount of prior volume where the dark matter interaction parameters become unconstrained. Qualitatively, we expect to see bounds derived from the posterior distribution on $\log_{10}(\sigma_0/\mathrm{cm}^2)$ become stronger for wider prior distributions.

To examine the impact of multiple parameters susceptible to prior-volume effects, we also run the analysis allowing $f_\chi$ to vary freely in addition to all the other model, cosmological, and nuisance parameters. We test four combinations of different flat prior ranges: $\log_{10}(f_\chi)\in [-2, 0]$ and $[-3.5, 0]$ with $\log_{10}(\sigma_0/\mathrm{cm^2})\in[-27,-19]$ and $[-31,-19]$. 

For each Bayesian analysis, we run four MCMC chains and determine chain sampling convergence using the Gelman-Rubin $R-1$ statistic. We require $R-1<0.005$ for all analyses that vary only two of the new model parameters. 
Given the higher computational cost of varying an additional parameter, particularly with prior-volume effects, we apply an adequate but less stringent convergence criteria of $R-1<0.01$ for our analyses that also vary $f_\chi$. In our presented results, we discard the first 30\% of the points in the chain files as ``burn-in.''

For the frequentist analyses, we obtain profile likelihoods using \texttt{Procoli}\footnote{\url{https://github.com/tkarwal/procoli}} \citep{Karwal_2024_Procoli}, a public profile-likelihood Python package that uses a simulated-annealing optimizer with \texttt{MontePython} \citep{Brinckmann_2018_MontePython}, in conjunction with any modified version of \texttt{CLASS}. 
To compare with our 2D marginalized posterior constraints, we report the 2D profile-likelihood constraints on the cross section and mass for the same fixed values of $\log_{10}(f_\chi)$ used in our MCMC analyses. 
With the fraction fixed, our parameters of interest are $\bm \mu = (\sigma_0, m_\chi)$, and our cosmological and nuisance parameters are $\bm\nu= \{ \ln(10^{10} A_\mathrm{s}),n_\mathrm{s},100\theta_\mathrm{s},\Omega_\mathrm{b} h^2, \Omega_\mathrm{c} h^2,\tau_\mathrm{reio},A_\mathrm{\textrm{Planck}} \}$.

We first find the global MLE $\mathcal{L}(\bm d|\bm{\hat\mu}, \bm{\hat\nu})$---i.e., the denominator in Eq.~\eqref{eqn:profilelikelihood}---where $\bm{\hat\mu}$ and $\bm{\hat\nu}$ are the best-fit parameter values that maximize the likelihood. 
Then, we fix the values of $\bm \mu = (\sigma_0, m_\chi)$ and optimize over all other parameters $\bm\nu$ to maximize the conditional likelihood, obtaining the numerator of Eq.~\eqref{eqn:profilelikelihood}. 
We fix $\log_{10}(m_\chi/\mathrm{GeV})$ to integer values from -5 to 3. 
In the cross section, we step sequentially away from the global MLE value of $\log_{10}(\sigma_0/\mathrm{cm^2})$ with step size $\pm0.2$ 
to balance resolution with efficiency.
Finally, with $\bm \mu = (\sigma_0, m_\chi)$, we have a two-dimensional profile likelihood, making $\Delta\chi^2=6.18$ the cutoff value for a 95.45\% confidence region.

Generating mock CMB data for each parameter value of interest is prohibitively computationally expensive, so we use the graphical method for constructing our confidence intervals. Requiring a non-negative value for the cross section places our limits near a physical boundary, and we consider the boundary-corrected graphical method in Appendix~\ref{app:Feldman-Cousins}. We find that our limits are far enough from the boundary to use the regular graphical method to obtain conservative upper limits---any modification from the boundary correction would result in slightly tighter constraints, though we show in Appendix~\ref{app:Feldman-Cousins} that the difference would not be significant enough to qualitatively change the comparison to the Bayesian constraints.

\begin{figure}
    \centering
    \includegraphics[width=\columnwidth]{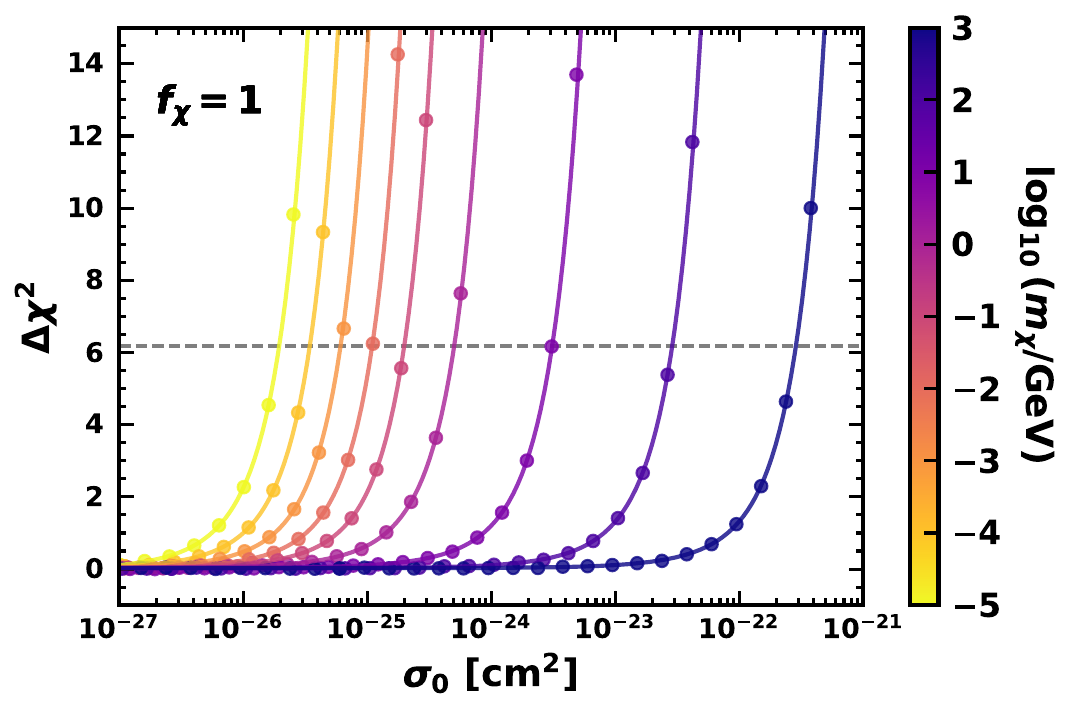}
    \caption{Profile likelihoods of the dark matter-proton scattering cross section $\sigma_0$ for different fixed dark matter particle masses $m_\chi$, as indicated by the color bar, for a fixed interaction fraction $f_\chi=1$. Points indicate computations of the profile likelihood, while the lines show the corresponding Gaussian fit. 
    The dashed horizontal line shows the $\Delta\chi^2=6.18$ cutoff value for a 95.45\% C.L., used to determine the upper limits on the cross section.
    }
    \label{fig:ProfileLikelihoods}
\end{figure}

Figure~\ref{fig:ProfileLikelihoods} shows the profile likelihood $\Delta\chi^2(\sigma_0)$ of the cross section for a fixed interaction fraction $f_\chi=1$ and different dark matter particle masses. The points indicate the values of the profile likelihood computed with \texttt{Procoli}. Taking our profile likelihoods in the form $e^{-\Delta\chi^2(\sigma_0)/2}$, we fit a Gaussian for linearly-spaced $\sigma_0$ values, also extrapolating into the unphysical region of negative cross section. 
We use the corresponding best-fit parabolas for $\Delta\chi^2$, shown as solid lines in Fig.~\ref{fig:ProfileLikelihoods}, to obtain constraints on the cross section.

The dashed horizontal line in Fig.~\ref{fig:ProfileLikelihoods} at $\Delta\chi^2=6.18$ indicates the cutoff value for a 95.45\% C.L. for a two-dimensional profile likelihood. 
Using the standard graphical method, the reported upper limit on the cross section is the value of $\sigma_0$ at which the profile likelihood $\Delta\chi^2(\sigma_0)$ intersects with $\Delta\chi^2=6.18$. 
The same analysis is used to obtain upper limits on the cross section as a function of mass for different fixed interaction fractions. The profile likelihoods for other values of $f_\chi$ are provided in Appendix~\ref{app:GaussProfLkl}.

\section{Results}\label{sec:Results}

We present CMB constraints for velocity-independent scattering between protons and a fraction of the total dark matter density. We compare our constraints using the profile-likelihood method to those we obtain using standard MCMC analyses. 
Finally, we test different priors in our MCMC analyses to demonstrate how constraints shift for parameters susceptible to prior-volume effects.

\subsection{Profile-likelihood constraints}

\begin{figure*}
    \centering 
    \includegraphics[width=\columnwidth]{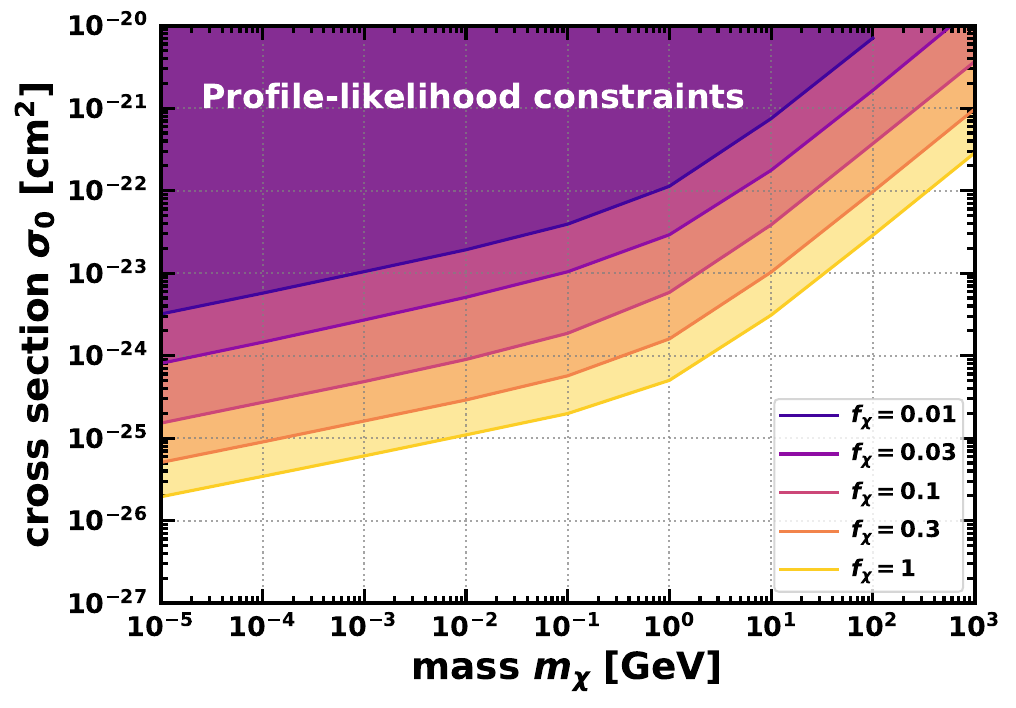}
    \includegraphics[width=\columnwidth]{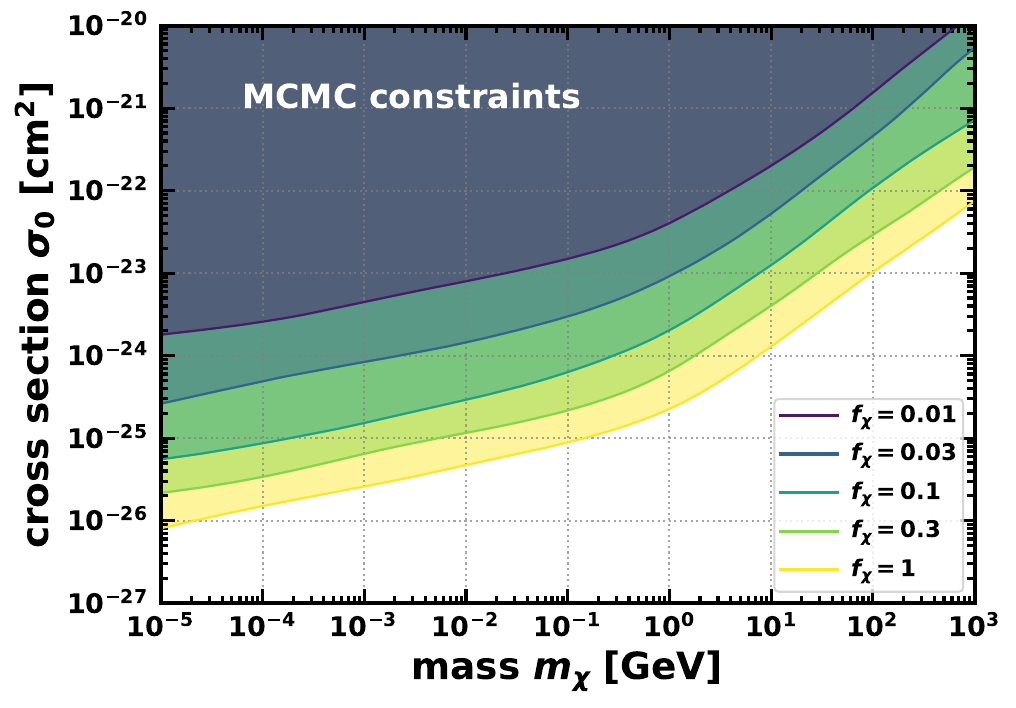}
    \caption{
    \textit{Left:} Profile-likelihood constraints on the cross section as a function of mass for the interaction fractions $f_\chi$ indicated in the legend. Shaded regions show the area of parameter space that is disfavored  at a 95.45\% confidence level. 
    \textit{Right:} Same as the left panel, but showing Bayesian constraints from our MCMC analyses. The edges of the excluded shaded regions correspond to the 95.45\% credible region contours of the 2D marginal posteriors. For each MCMC analysis, we adjust the priors to accommodate larger cross sections for small interaction fractions as outlined in \cref{tab:f_sigma_priors}. 
    }
    \label{fig:constraints}
\end{figure*}

We present the profile-likelihood constraints on the dark matter-proton momentum-transfer cross section $\sigma_0$ as a function of the dark matter particle mass $m_\chi$ for different interaction fractions $f_\chi$ in the left panel of Fig.~\ref{fig:constraints}.
Curves indicate the upper limits obtained on $\sigma_0$ where its profile likelihood intersects $\Delta\chi^2$=6.18, as described in Sec.~\ref{sec:Methods}. The shaded regions where $\Delta\chi^2>6.18$ are excluded at a 95.45\% confidence level. 

The constraints capture the expected relationship between the cross section and mass. 
For high dark matter masses $m_\chi\gg m_\mathrm{p}$, the cross section and mass become degenerate and reduce to the single parameter $\sigma_0 / m_\chi$, resulting in a constraint that scales as $\sigma_0 \propto m_\chi$ (e.g., Refs.~\cite{Dvorkin_2013_constraining,GluscevicBoddy_2018_constraints}).
For low dark matter masses $m_\chi\ll m_\mathrm{p}$, the temperature evolution of dark matter plays a nontrivial role in setting the shape of the constraint (e.g., Refs.~\cite{GluscevicBoddy_2018_constraints,Boddy_2018_new,Nadler_2019_Constraints,Nguyen_2021_observational}).
The transition between these two limits occurs for $m_\chi \approx m_\mathrm{p}$.

\subsection{Comparison to MCMC constraints}

The right panel of Fig.~\ref{fig:constraints} shows the same constraints, but for the Bayesian MCMC method. The curves show the 95.45\% credible region contours of the 2D marginal posterior distributions, and the shaded regions indicate the excluded regions of parameter space. The profile-likelihood and MCMC constraints broadly agree: no preference is found for dark matter-proton interactions, and the shapes of the upper limits are consistent with expectations for both methods.

\begin{figure}
    \centering
    \includegraphics[width=\columnwidth]{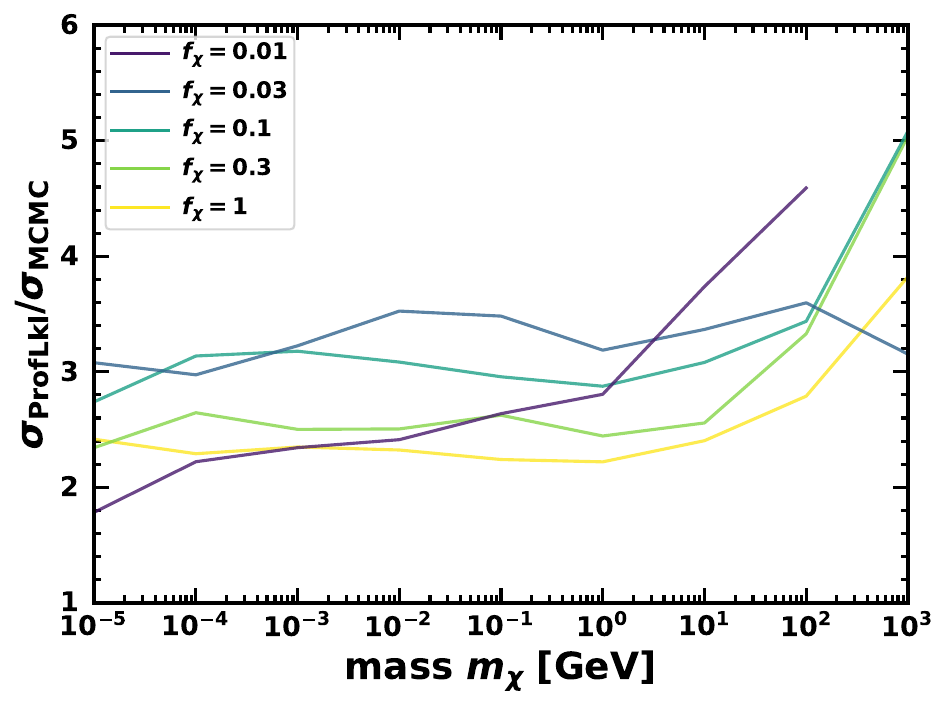}
    \caption{The ratio of the profile-likelihood upper limits on the interaction cross section $\sigma_{\rm ProfLkl}$ to the MCMC limits $\sigma_{\rm MCMC}$ for fixed interaction fractions $f_\chi$ as indicated by different colors. The MCMC posteriors are more constraining than the profile likelihoods across the full mass range for every choice of fraction.
    }\label{fig:FreqBayesRatio}
\end{figure}

However, beyond this general agreement, the MCMC limits are systematically stronger than the profile-likelihood constraints by a factor of two or more across the entire mass range for all fractions, i.e., the profiles permit larger cross sections. In Fig.~\ref{fig:FreqBayesRatio}, we show the ratio of the profile-likelihood to the MCMC constraints on the cross section as a function of mass for the different interaction fractions. 
Note that the purple curve terminates early because $f_\chi = 0.01$ crosses $m_\chi = 10^3$ GeV beyond the range explored in $\sigma_0$ (see \cref{fig:constraints,fig:ProfileLikelihoods_f}). 
The discrepancy between $\sigma_{\rm ProfLkl}$ and $\sigma_{\rm MCMC}$ is consistent with our expectations for prior-volume effects, which we explore in more detail in the next section.

\subsection{Impact of priors}
\label{subsec:priors}

\subsubsection{Varying cross section priors}

\begin{figure}
    \centering
    \includegraphics[width=\columnwidth]{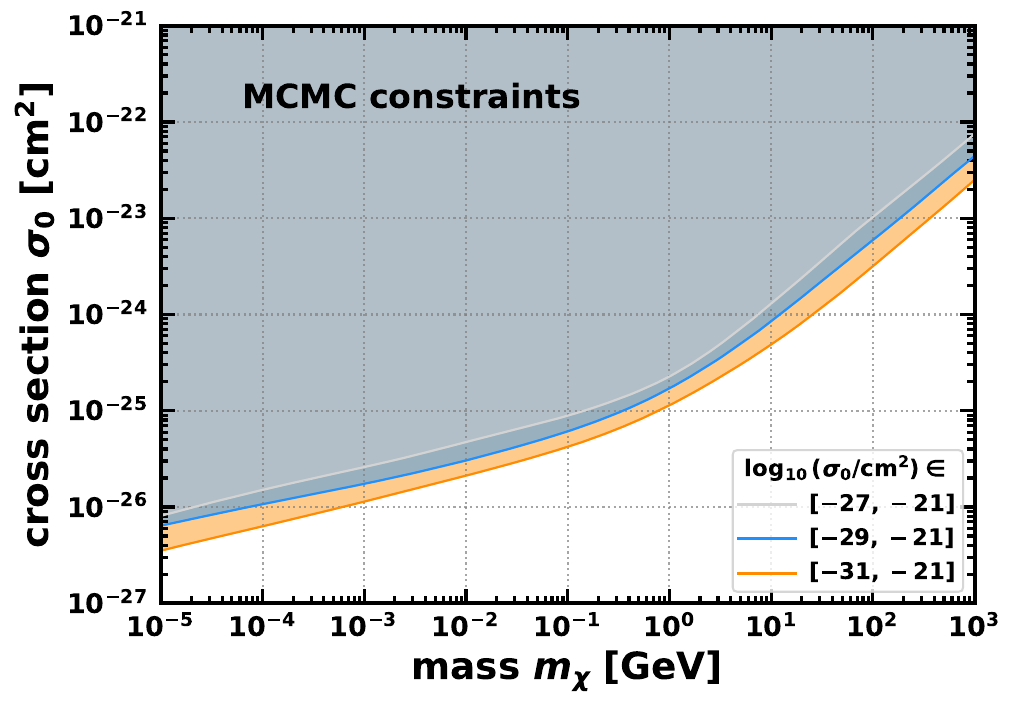}
    \includegraphics[width=\columnwidth]{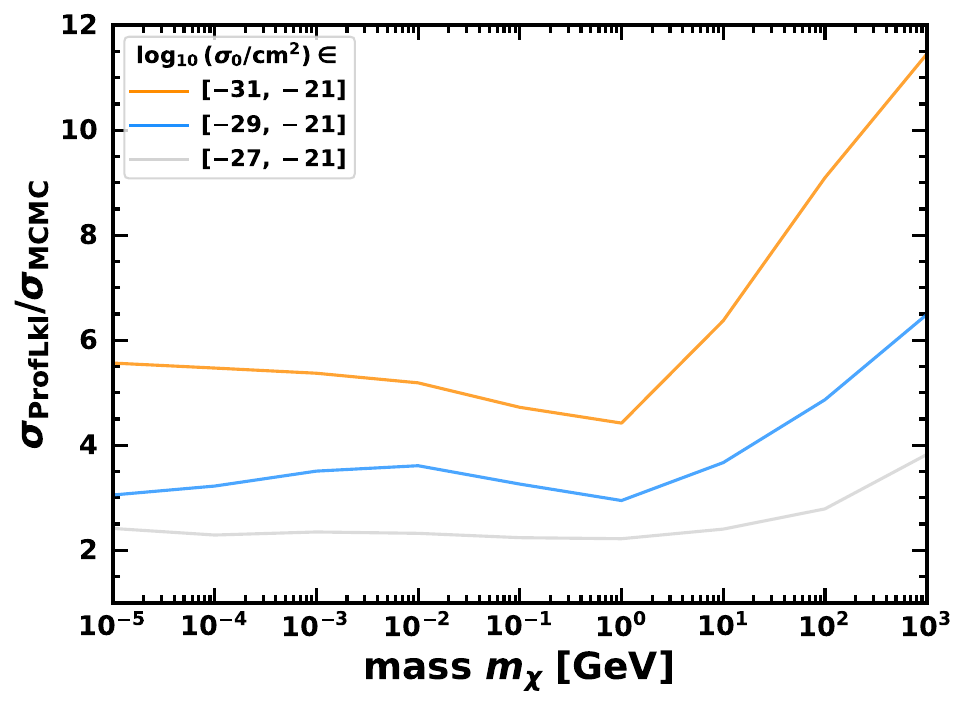}
    \caption{\textit{Upper:} Bayesian constraints obtained for three different priors on the cross section $\sigma_0$ as indicated, all for fixed interaction fraction $f_\chi=1$. The shaded regions indicate the parameter space excluded at a 95.45\% credible level.
    \textit{Lower:} The same plot as Fig.~\ref{fig:FreqBayesRatio} but for different priors on the cross section as indicated, all for fixed interaction fraction $f_\chi=1$. 
    The Bayesian posteriors are clearly influenced by the choice of prior: as the prior on $\log_{10}(\sigma_0/\mathrm{cm^2})$ widens, the Bayesian limits become more constraining since the posterior is pulled towards the larger prior volume that is indistinguishable from $\Lambda$CDM.
    }
     \label{fig:PriorVolumeEffects}
\end{figure}

To investigate how the choice of prior on $\log_{10}(\sigma_0/\mathrm{cm}^2)$ affects the MCMC constraints, we fix the fraction to $f_\chi=1$ and fix the upper bound of the prior on $\log_{10}(\sigma_0/\mathrm{cm}^2)$. We then run three analyses varying the lower bound on the prior. 
The top panel in Fig.~\ref{fig:PriorVolumeEffects} shows the resulting MCMC constraints. 

The posteriors show a clear dependence on the choice of prior: as expected, as the prior range widens to allow lower cross sections, the MCMCs preferentially sample the asymptotic $\Lambda$CDM limit, and the posterior shifts towards lower values of the cross section.
Extending the lower bound of the prior range in $\log_{10}(\sigma_0/\mathrm{cm^2})$ by four orders of magnitude leads to limits on $\sigma_0$ that are at least two or three times as constraining. We emphasize that the chains are fully converged, according to the criteria specified in Sec.~\ref{sec:Methods}. 
The different posteriors are the result of the larger prior volume in the $\Lambda$CDM limit where the data can no longer distinguish between the models.

We compare the constraints obtained using different priors on the cross section to the $f_\chi=1$ profile-likelihood constraints in the lower plot of Fig.~\ref{fig:PriorVolumeEffects}. 
As the prior range widens to include lower values of the cross section, the limits become more stringent and deviate more from the profile-likelihood limits. Additionally, the posteriors do not fully capture the $\sigma_0 \propto m_\chi$ relationship as $m_\chi \gg m_\mathrm{p}$ because sampling a larger prior that includes more of the $\Lambda$CDM limit pulls the posterior towards smaller values of the cross section and slightly flattens the slope of the constraint at higher masses.

This downward shift in the marginalized posterior in the $\sigma_0$-$m_\chi$ plane towards lower values of $\sigma_0$ should continue indefinitely as more prior volume is included at low $\log_{10}(\sigma_0/\mathrm{cm^2})$.
In logspace, there is infinite volume as $\sigma_0 \rightarrow 0$, and an arbitrarily broad prior can be chosen, encompassing more of this space. 
Moreover, the likelihood flattens at low $\log_{10}(\sigma_0/\mathrm{cm^2})$ as seen in \cref{fig:ProfileLikelihoods}, and the global MLE is in this $\sigma_0 \rightarrow 0$ limit. 
Since Bayesian posteriors are a product of the prior and the likelihood, expanding the uniform prior range to include more of this region with a flat likelihood simply shifts the posterior volume towards $\sigma_0 \rightarrow 0$, resulting in increasingly stronger constraints.
Critically, we note that this shifting of constraints does not happen for the prior-independent profile-likelihood confidence limits. Adding more points at lower $\log_{10}(\sigma_0/\mathrm{cm^2})$ in \cref{fig:ProfileLikelihoods} has no impact on where the $\Delta \chi^2$ threshold is crossed and therefore no impact on the $95.45\%$ C.L.s.

\subsubsection{Varying both cross section and fraction priors}

\begin{figure*}
    \centering
    \includegraphics[width=\textwidth]{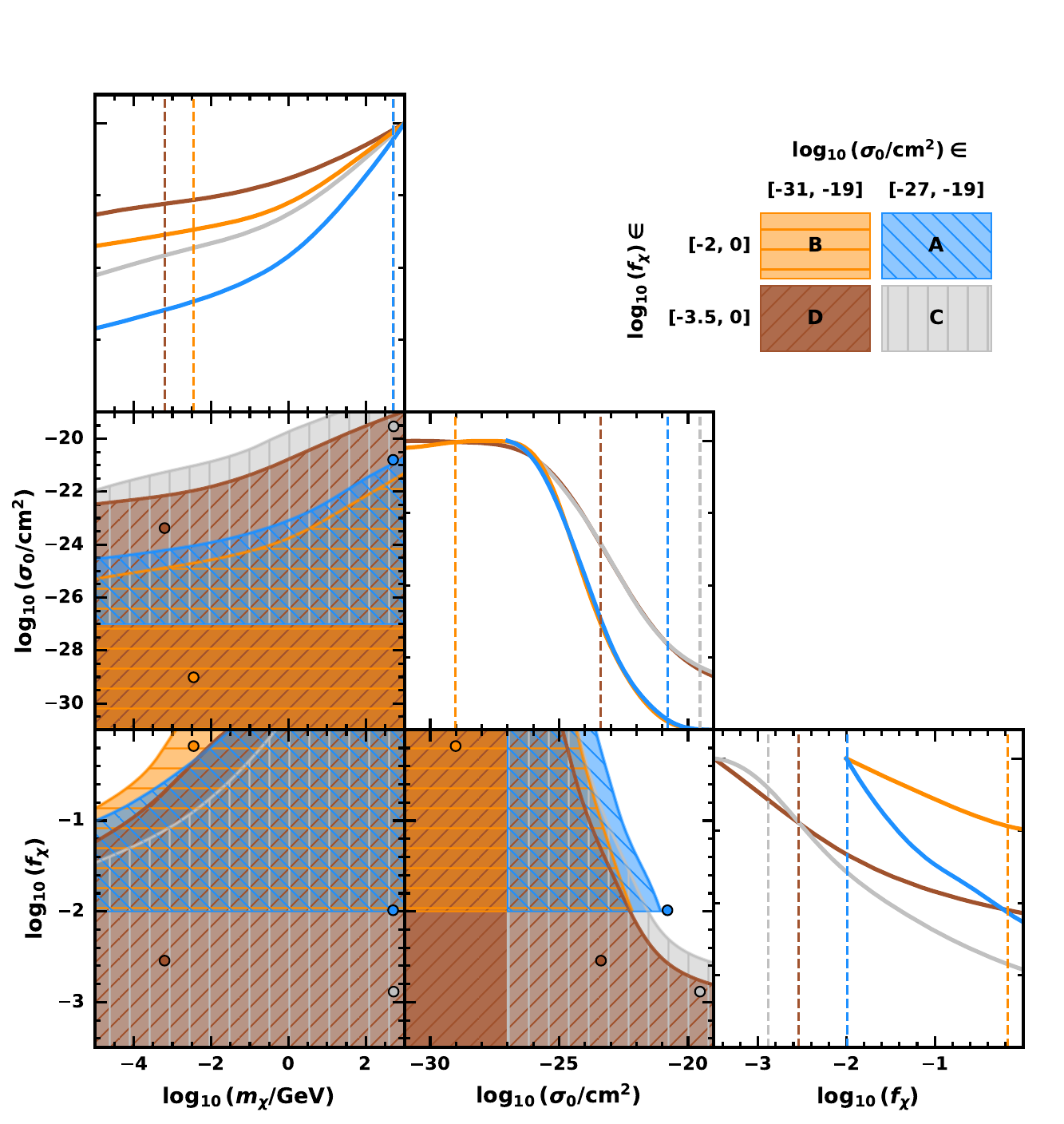}
    \caption{Marginal posteriors for the three dark matter-proton scattering parameters for four combinations of priors on the cross section $\sigma_0$ and interaction fraction $f_\chi$ indicated by different colors. 
    Here, shaded regions mark the favored $95.45\%$ credible regions of the parameter space. Expanding the range in either prior to include smaller values causes the posteriors to shift towards the $\Lambda$CDM limit, resulting in tighter constraints on that parameter. Marginalizing over a broader prior volume in the $\Lambda$CDM limit for the cross section results in looser constraints on the fraction, and vice versa. 
    The colored points (with corresponding vertical lines in the one-dimensional plots) show the global maximum likelihood estimates calculated within the corresponding parameter ranges for each prior choice, also recorded in \cref{tab:f_sigma_priors}. We find all the MLEs in a region of parameter space indistinguishable from $\Lambda$CDM with values consistent with the MLE for $\Lambda$CDM.
    }
    \label{fig:3D_Posteriors}
\end{figure*}

To understand the full impact of the priors in the MCMC analysis, we simultaneously vary all three new model parameters, including the interaction fraction $f_\chi$. In cases A-D shown in \cref{tab:2D_priors}, we consider a total of four combinations of priors for $\log_{10}(f_\chi)$ and $\log_{10}(\sigma_0/\mathrm{cm^2})$, the two parameters causing prior-volume effects.
Figure~\ref{fig:3D_Posteriors} shows the marginalized posteriors for the mass, cross section, and fraction. Note that shaded regions here represent parameter space allowed at a 95.45\% credible level. The colored points show the global MLEs calculated within the corresponding parameter ranges. 

\begin{table}[]
    \centering
    \begin{tabular}{c|c|c|c}
        Prior label & \shortstack{Prior on \\$\log_{10}(f_\chi)$} & \shortstack{Prior on \\ $\log_{10}(\sigma_0/\mathrm{cm^2})$} & \shortstack{Global MLE \\ $-\ln{\mathcal{L}(\bm d|\bm{\hat\mu}, \bm{\hat\nu}})$} \\
        \hline \hline
        A & $[-2, 0]$  & $[-27,-19]$ & $506.44$ \\
        B & $[-2, 0]$  & $[-31,-19]$ & $506.42$ \\
        C & $[-3.5,0]$ & $[-27,-19]$ & $506.35$ \\
        D & $[-3.5,0]$ & $[-31,-19]$ & $506.26$ \\
    \end{tabular}
    \caption{Prior combinations for the two new model parameters that lead to prior-volume effects $\log_{10}(f_\chi)$ and $\log_{10}(\sigma_0/\mathrm{cm^2})$. The upper bounds are fixed for both parameters. For each combination, we also show the global MLE, effectively identical to the $\Lambda$CDM global MLE $ = 506.42$. }
    \label{tab:2D_priors}
\end{table}

Although the calculated global MLEs $\mathcal{L}(\bm d|\bm{\hat\mu}, \bm{\hat\nu})$ occur for different best-fit values $\bm{\hat\mu}$, they all occur for essentially the same minimum value of the negative log-likelihood, as shown in \cref{tab:2D_priors}.
All these MLE points fall in the $\Lambda$CDM limit of the parameter space, where either the cross section is very small (B) or the fraction is very small (A, C, D), as shown in \cref{fig:3D_Posteriors}. In either of these cases, the other new parameters, such as the mass, have a negligible effect on the likelihood. 
Indeed, we find $-\ln{\mathcal{L}(\bm d|\bm{\hat\mu}, \bm{\hat\nu}}) = 506.42$ when fitting the $\Lambda$CDM model, indicating that the calculated global MLEs are in a region of parameter space where the likelihood is flat and the data cannot distinguish between the models. 
The global MLE recovered is therefore to some extent arbitrary---as long as they are firmly within the $\Lambda$CDM limit, the values of the new-model parameters that maximize the likelihood are inconsequential. Even if a different set of values is recovered for the global MLE, as long as it is within this $\Lambda$CDM limit, the profile-likelihood limits would not be affected.

We find that the posteriors depend on the choice of priors and shift across cases A-D. 
The 1D marginal posterior distributions of the cross section and fraction in Fig.~\ref{fig:3D_Posteriors} show that extending the lower bound of either prior pulls the marginal posterior of that parameter towards the $\Lambda$CDM limit, resulting in tighter constraints on the model. We see the same effect in the 2D marginal posteriors. 
The 2D marginal posterior distribution of the cross section and mass shows that broadening the prior range for $\log_{10}(\sigma_0/\mathrm{cm^2})$ to include smaller cross sections (A$\rightarrow$B and C$\rightarrow$D) results in tighter, more stringent constraints on $\sigma_0$, consistent with the behavior seen in the upper panel of Fig.~\ref{fig:PriorVolumeEffects}. 
Similarly, the 2D marginal posterior of the fraction and mass in the lower left plot of Fig.~\ref{fig:3D_Posteriors} shows that widening the prior range for $\log_{10}(f_\chi)$ to include smaller fractions (A$\rightarrow$C and B$\rightarrow$D) results in more stringent constraints on $\log_{10}(f_\chi)$. This is the same behavior discussed in the previous subsection and visualized in \cref{fig:PriorVolumeEffects}, where the MCMC simply samples the prior in the asymptotic $\Lambda$CDM limit, leading to tighter constraints when the prior contains a larger volume indistinguishable from $\Lambda$CDM.

We observe a different effect on the constraints for the cross section and fraction when the prior on the other parameter is widened. Specifically, marginalizing over a broader prior volume in the $\Lambda$CDM limit for one parameter results in looser constraints on the other.
For example, in the 1D marginal posterior of $\log_{10}(\sigma_0/\mathrm{cm^2})$, C and D show higher probability densities for larger cross sections than A and B because they are marginalizing over a wider range of very small $\log_{10}(f_\chi)$ values. 
Smaller values of $\log_{10}(f_\chi)$ permit larger values of $\log_{10}(\sigma_0/\mathrm{cm^2})$ while remaining consistent with the $\Lambda$CDM limit; thus, marginalizing over a wider range of interaction fractions loosens the constraints on the cross section, and vice versa. 

The 2D marginal posterior distribution of the cross section and mass also shows that broadening the prior range on $\log_{10}(f_\chi)$ to include smaller fractions (A$\rightarrow$C and B$\rightarrow$D) results in significantly less stringent constraints on $\log_{10}(\sigma_0/\mathrm{cm^2})$.
In the same way, the 2D marginal posterior of the fraction and mass shows that widening the prior range on $\log_{10}(\sigma_0/\mathrm{cm^2})$ to include smaller cross sections (A$\rightarrow$B and C$\rightarrow$D) loosens the constraints on $\log_{10}(f_\chi)$ because of the marginalization over a wider range of low cross sections. 
Altogether, we find that the choice in the prior for one parameter changes the constraints on the other new model parameter---an impact that can only be fully appreciated in an analysis that simultaneously varies all three of the interacting dark matter model parameters.

We also highlight that the 2D marginal posterior distribution of the fraction and mass (lower left corner of Fig.~\ref{fig:3D_Posteriors}) shows a region for low masses in which 10-100\% interaction is entirely excluded from the 95.45\% credible region. This constraint is set by the prior, not the likelihood. 
As the orange point shows, the likelihood in this region of parameter space can be consistent with the MLE for standard $\Lambda$CDM, provided that the cross section is small enough. The posterior constraints only exclude this region because prior-volume effects in the cross section and fraction cause 95.45\% of the parameter chain to fall in the inflated region of parameter space where the new parameters have a negligible effect on the model predictions.

\section{Conclusions}\label{sec:Conclusions}

Parameter inference in cosmology generally relies on Bayesian statistics, but obtaining constraints on models such as interacting dark matter under this statistical framework are complicated by prior-volume effects. 
For the dark matter-baryon scattering model, as the interaction fraction or cross section go to zero, the other new parameters of the model become unconstrained. 
A large prior volume that includes the $\Lambda$CDM limit can result in artificially tighter constraints on the model parameters from Bayesian posteriors. This motivates the use of frequentist profile-likelihoods constraints that are independent of priors.

Using CMB data, we present the first profile-likelihood constraints on the dark matter-proton scattering model that consider fractional cases (Fig.~\ref{fig:constraints}). For comparison, we also perform a Bayesian MCMC analysis and find general agreement in the shape of the upper limits. However, we find that constraints from the Bayesian marginal posterior distributions of the cross section and mass place limits that are stronger by a factor of two or more compared to the profile likelihoods (Fig.~\ref{fig:FreqBayesRatio}). Given these discrepancies, we recommend caution when interpreting the results of Bayesian constraints on this and similar models.

By testing different priors on the cross section and interaction fraction in our MCMC analyses, we demonstrate that the posteriors are indeed affected by prior choices. 
Broadening the lower bound of the prior on the cross section results in tighter constraints on the cross section (\cref{fig:PriorVolumeEffects}). 
We expect this prior-dependent shift in constraints to continue indefinitely for smaller lower bounds on the cross section prior, as the likelihood flattens in this direction (\cref{fig:ProfileLikelihoods}) and Bayesian posteriors simply recover the flat prior, effectively gaining more volume in the $\Lambda$CDM limit region.

Prior-volume effects are exacerbated when simultaneously varying the interaction fraction, cross section, and mass (Fig.~\ref{fig:3D_Posteriors}). 
Wider priors that include smaller values of the cross section or fraction cause the posteriors of those parameters to shift towards the smaller values, to a region of parameter space that is indistinguishable from $\Lambda$CDM and thus where the other new model parameters can take on any value without changing the observables. Additionally, marginalizing over a broader prior volume in the $\Lambda$CDM limit for the cross section results in looser constraints on the fraction, and vice versa. Therefore, the choice of prior for one parameter affects constraints for multiple parameters.

One potential indication of the existence of prior-volume effects is a mismatch between the mean of Bayesian posteriors and the maximum likelihood for a subset of parameters. 
For example, in the case of the early dark energy (EDE) model that alleviates the Hubble tension \cite{Karwal:2016vyq,Poulin_2018_Early,Poulin:2023lkg,Kamionkowski:2022pkx}, varying the fraction of EDE introduces prior-volume effects into constraints on the model, similar to how varying $f_\chi$ does in our model. 
Its Bayesian posteriors peak at $0$ (the $\Lambda$CDM limit) and only result in upper limits \cite{AtacamaCosmologyTelescope_2025,SPT-3G:2025vyw}, yet the MLE lies beyond $2\sigma>0$, a mismatch that indicates the influence of prior-volume effects on the posteriors \cite{Poulin:2023lkg,Karwal_2024_Procoli}. 
In contrast, for our model, the means and maximum likelihoods agree and occur well within the $\Lambda$CDM limit, demonstrating that this quick test to rule out prior-volume effects in the posteriors is not sufficient.
Although prior volume does not bias the means recovered from Bayesian posteriors, it does shift the limits. 
This serves as a cautionary tale on thoroughly understanding the influence of prior choices on models when extracting constraints via Bayesian techniques.

Although CMB observations do not provide the most stringent constraints for velocity-independent dark matter-baryon scattering, they are a clean and well-studied probe, suitable to use as a basis for investigating prior-volume effects. 
Regardless, we expect that our qualitative conclusions also apply to other cosmological probes and similarly parametrized models. Prior-volume issues arise in this model for any MCMC analysis, barring a dataset exhibiting strong evidence in favor of dark matter-baryon scattering. More generally, unless strong theoretical arguments warrant the use of specific priors, we recommend using both MCMC and profile-likelihood approaches to place upper limits on new-physics models with parameters susceptible to prior-volume effects. Doing so helps produce a clearer understanding of cosmological limits imposed by the data as well as the influence of priors for the model in consideration.

%%%%%%%%%%%%%%%%%%%%%%%%%%%%%%%%%%%%%%%%%%%%%%%%%%%%%%%%%%%%%%%%%%%%%%%%%%%%%%%
\begin{acknowledgments}
The ideas for this paper came together at the Understanding Cosmological Observations 2023 conference in Benasque, Spain for which we thank the organizers. 
MCS acknowledges support from the NSF Graduate Research Fellowship Program under Grant No. DGE 2137420.
TK is supported by the Kavli Institute for Cosmological Physics at the University of Chicago through an endowment from the Kavli Foundation. 
JLB acknowledges funding from the project UC-LIME (PID2022-140670NA-I00), financed by MCIN/AEI/ 10.13039/501100011033/FEDER, UE.
KB acknowledges support from the NSF under Grant No. PHY-2413016. 
This work used Stampede3 at the Texas Advanced Computing Center (TACC) through allocation PHY240163 from the Advanced Cyberinfrastructure Coordination Ecosystem: Services \& Support (ACCESS) program, which is supported by U.S. National Science Foundation grants \#2138259, \#2138286, \#2138307, \#2137603, and \#2138296.
We thank the developers of the Python packages used in preparing this paper: \texttt{NumPy} \citep{Harris_2020_NumPy}, \texttt{SciPy} \citep{Virtanen_2020_SciPy}, \texttt{GetDist} \citep{Lewis_2019_GetDist} and \texttt{Matplotlib} \citep{Hunter_2007_Matplotlib}. 

\end{acknowledgments}

%%%%%%%%%%%%%%%%%%%%%%%%%%%%%%%%%%%%%%%%%%%%%%%%%%%%%%%%%%%%%%%%%%%%%%%%%%%%%%%

\appendix

\section{Boundary-corrected graphical method}\label{app:Feldman-Cousins}

Constructing a Frequentist confidence interval with guaranteed correct coverage would require simulating hundreds of mock CMB observations for each set of candidate true values of all the parameters being considered. Due to statistical fluctuations, the many simulations would produce a spread of best-fit values for each candidate true value. 
The acceptance region for each candidate true value is the set of best-fit values, admitted in order of the log-likelihood ratio statistic (Eq.~\ref{eqn:profilelikelihood}), that contains 95.45\% of the simulations. The 95.45\% confidence interval would be the range of candidate true values whose acceptance regions include our observed best-fit value. 

Simulating mocks is computationally expensive and can be avoided in the asymptotic regime, where the confidence interval is obtained directly from the shape of the profile likelihood using the graphical method. In the asymptotic limit of a large data set, the log-likelihood ratio statistic follows a $\chi^2$ distribution according to Wilks' Theorem \citep{Wilks_1938_Large-Sample}. The $\Delta\chi^2$ cutoff for a 95.45\% profile-likelihood confidence interval is defined as the 95.45th percentile of the $\chi^2$ distribution with $k$ degrees of freedom.
For example, this confidence interval corresponds to a cut of $\Delta\chi^2=4$ or $\Delta\chi^2=6.18$ for a $\chi^2$ distribution with one or two degrees of freedom, respectively, which we denote as $\chi^2_1$ or $\chi^2_2$.

Reference~\citep{Herold_2024_profile} generated mock \textit{Planck}-lite spectra and verified that the asymptotic limit and Wilks’ theorem hold for the $\Lambda$CDM parameters, in which case the graphical method for obtaining confidence intervals yields correct coverage. 

However, Wilks’ theorem no longer holds when a parameter $\bm{\mu}$ is near a physical boundary, e.g. $\bm{\mu} > 0$, as its log-likelihood ratio statistic is no longer $\chi^2$-distributed. In this case, the usual graphical method for constructing Frequentist confidence intervals does not guarantee correct coverage. 

The boundary correction can be handled by using the Feldman-Cousins construction \citep{Feldman_1997_Unified}, which changes the ordering rule for building the confidence region by modifying Eq.~\eqref{eqn:profilelikelihood} so that the MLE in the denominator is restricted to the physically allowed region $\bm{\mu}\geq0$. 
Accordingly, proximity to a physical boundary leads to changes in the asymptotic theory \citep{Cowan_2010_Asymptotic,Salucci:2025vem} and a modified graphical construction. 

We examine the modified graphical construction for one free parameter and two free parameters. 

\subsection{1D boundary-corrected graphical method}

Because the best-fit cross section cannot be negative, repeated experiments for a true value near the boundary would find that the best-fit values pile up at the boundary. This reshapes the distribution of the log-likelihood test statistic, which no longer follows $\chi^2_1$, and so integrating the distribution reaches 95.45\% at a lower $\Delta\chi^2$ threshold. 
At the boundary $\bm{\mu} = 0$, the modified relationship for the log-likelihood ratio statistic is a half-parabola that is parabolic for $\bm{\mu}>0$ and flat at zero for $\bm{\mu}<0$, since the constrained best fit is pinned at $\bm{\mu}=0$. 
According to Chernoff's theorem \citep{Chernoff_1954_Distribution}, this results in a half-$\chi^2$ distribution, $\frac{1}{2}\chi_0^2+\frac{1}{2}\chi^2_1$. The constrained best fit cannot move below the boundary, so those realizations pile up at a test-statistic value of zero (i.e., $\chi_0^2$ is a delta function). 
The other half follows $\chi^2_1$ because of the expected positive scatter from one free parameter. 
In this mixture, the weight of $\chi_0^2$ shrinks and the weight of $\chi^2_1$ grows as $\bm{\mu}$ moves away from the boundary. Far away from the boundary, the standard $\chi^2$ distribution with one degree of freedom is recovered, and so the modification makes no difference for $\bm{\mu}\gg0$. 

The boundary-corrected confidence-interval construction method has been used to obtain profile-likelihood bounds on the cosmological sum of neutrino masses $M_\nu$, where $M_\nu$ is physically constrained to be non-negative \citep{Herold_2024_profile,Naredo-Tuero_2024_Critical,Chebat_2025_Cosmological}.  Reference~\citep{Herold_2024_profile} generated mock spectra to test $\Lambda$CDM+$M_\nu$ and found that the distribution of $M_\nu$ is consistent with a Gaussian near a physical boundary, which validates the use of the boundary-corrected graphical construction to obtain confidence intervals. 

\begin{figure}
    \centering
    \includegraphics[width=\columnwidth]{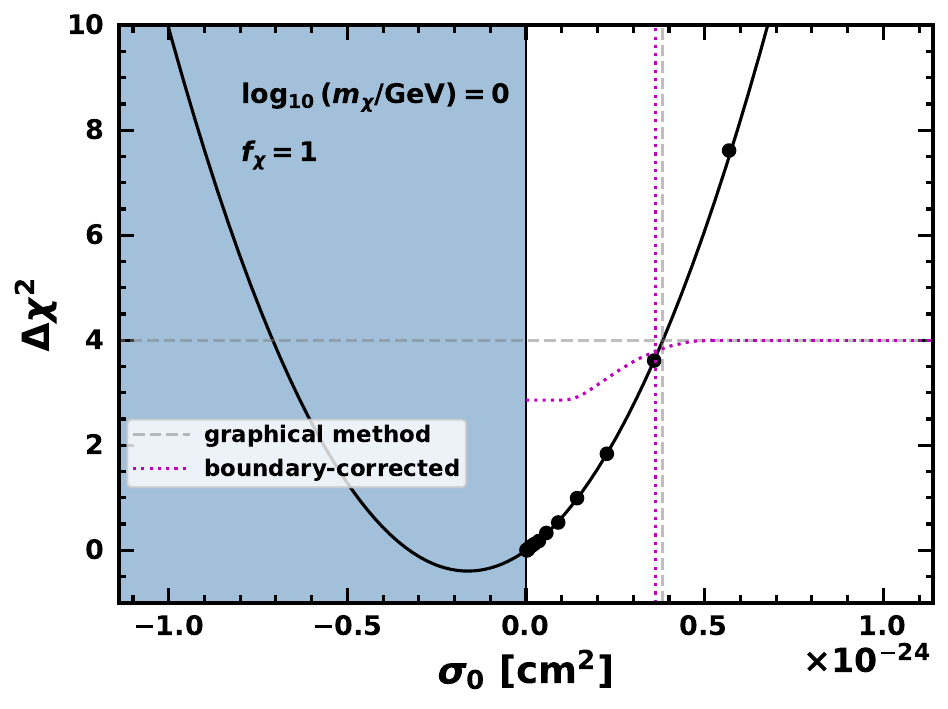}
    \caption{
    The cross section profile likelihood for the indicated mass and fraction shown by points, with the fit parabola in solid black. 
    The horizontal dashed gray line shows the $\Delta\chi^2=4$ for a one-dimensional profile likelihood using the graphical method, while the the dotted magenta line indicates the $\Delta\chi^2$ cutoff for the boundary-corrected graphical method (see Table III of Ref.~\citep{Herold_2024_profile}). 
    The vertical dashed gray line shows the 95.45\% confidence interval upper limit on $\sigma_0$ using the regular graphical method, and the vertical dotted magenta line shows the constraint on $\sigma_0$ using the boundary-corrected graphical method.
    The blue shaded region has unphysical $\sigma_0 < 0$. 
    }
    \label{fig:FC_m0_f1}
\end{figure}

In our model, we similarly require a non-negative value for $\sigma_0$, which places our constraints near a physical boundary. 
For the simplicity of only considering one free parameter, we first apply the boundary-correction to the 1D per-mass constraint on the cross section. In Fig.~\ref{fig:FC_m0_f1}, we plot the cross section profile likelihood for $m_\chi=1$ GeV and $f_\chi=1$ along with the $\Delta\chi^2$ cutoff for the boundary-corrected graphical method. We obtain the cutoff using Table III of Ref.~\citep{Herold_2024_profile}, which provides the correction for a Gaussian near a physical boundary for a 95\% C.L. as a function of the model parameter in units of the standard deviation. 
The solid line shows the fit parabola, extrapolating to negative cross sections.

\begin{figure}
    \centering
    \includegraphics[width=\columnwidth]{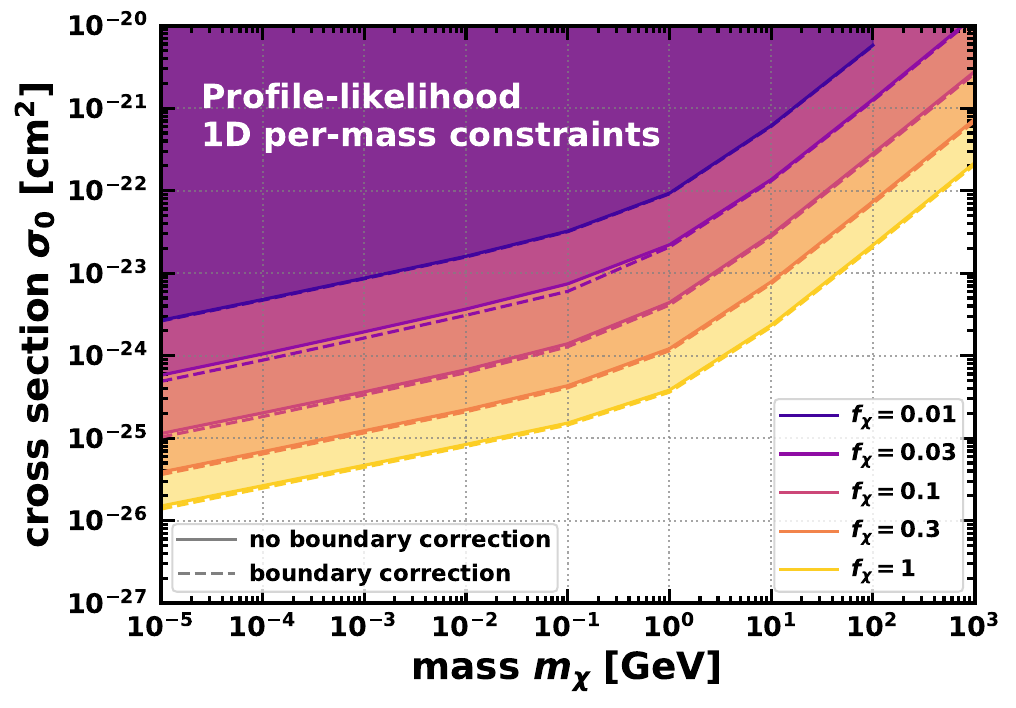}
    \includegraphics[width=\columnwidth]{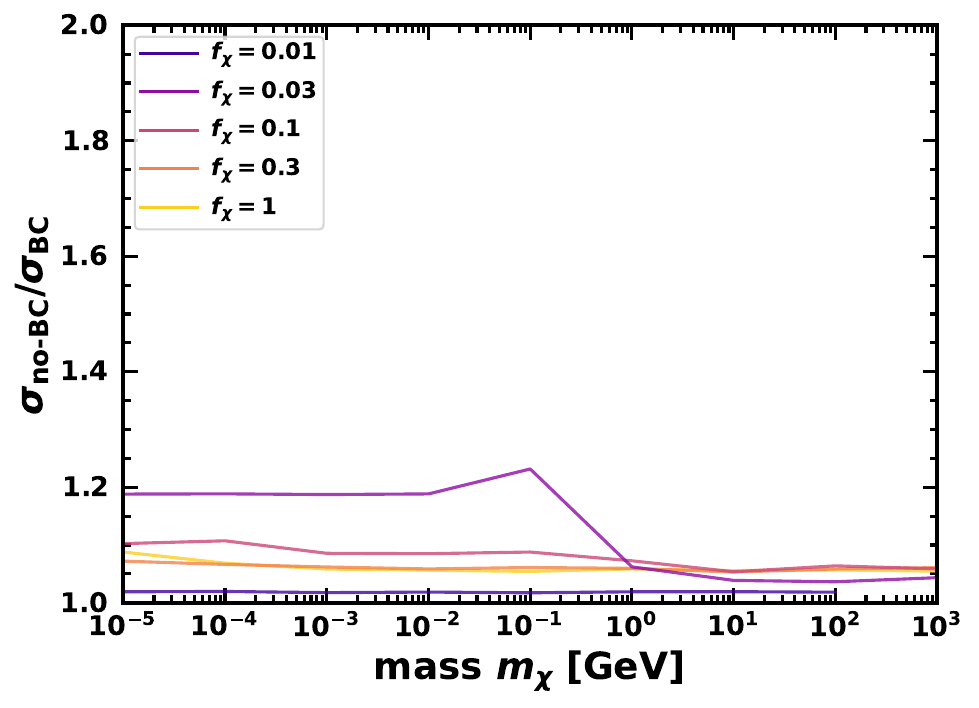}
    \caption{\textit{Upper:} One-dimensional profile likelihood constraints on the cross section at fixed masses. 
    The shaded regions indicate parameter space excluded from the 95.45\% confidence interval drawn using the graphical method (solid) and the boundary-corrected graphical method (dashed) for fixed interaction fractions indicated by color.  
    \textit{Lower:} The ratio comparing the graphical method and the boundary-corrected graphical method constraints on the cross section per fixed mass for fixed interaction fractions indicated by color. 
    }
     \label{fig:BoundaryCorrectedComparison}
\end{figure}

Using the boundary-corrected graphical method, we find the upper limit for the cross section at each fixed mass. 
The upper plot of Fig.~\ref{fig:BoundaryCorrectedComparison} shows the 1D per-mass profile-likelihood constraints on the cross section with and without using the boundary correction for the graphical method. The lower plot shows the ratio between the two methods. In most cases, the constraint is drawn far enough above the boundary (about two standard deviations) that the boundary correction is small.

\subsection{2D boundary-corrected graphical method}

The boundary correction is simplest to consider for one free parameter, in which case the distribution of the log-likelihood ratio statistic becomes a mixture of a delta function $\chi_0^2$, and a $\chi^2$ distribution  with one degree of freedom $\chi_1^2$. However, we present 2D constraints on the cross section and mass and therefore must consider the effect of two free parameters on the distribution of the log-likelihood ratio statistic.

A second free parameter would add one degree of freedom to each part of the mixture. At the boundary, if both parameters are Gaussian-distributed, the distribution of the log-likelihood ratio statistic would be $\frac{1}{2}\chi_1^2 + \frac{1}{2}\chi_2^2 $ \citep{Self_1987_Asymptotic}. Half of the time the cross section value scatters above the boundary, in which case there are two degrees of freedom, and half of the time the cross section is pinned at the boundary, in which case there is only the degree of freedom from the new free parameter. Far from the boundary, the distribution of the log-likelihood ratio statistic would again become a $\chi^2$ distribution with two degrees of freedom.

However, for a free parameter to contribute a full degree of freedom to the $\chi^2$ distribution of the log-likelihood ratio statistic, the profile likelihood in that parameter must be parabolic, i.e., changes to that parameter must degrade the quality of the fit in a well-defined quadratic manner. As we saw with our cross section, being near a physical boundary resulted in a profile likelihood that is a half-parabola, in which case the distribution of the log-likelihood ratio statistic is a half-$\chi^2$ mixture. 

The shape of the mass profile likelihood is more complicated. 
If the likelihood is flat for changes in a parameter, it will not contribute a clean additional degree of freedom to the $\chi^2$ distribution of the log-likelihood ratio statistic, which will affect the appropriate $\Delta\chi^2$ cut for a 95.45\% confidence region. In the regime of very small cross section, the mass profile likelihood becomes flat. At the boundary where the cross section is zero, the mass has no impact on the likelihood at all.

\begin{figure}
    \centering
    \includegraphics[width=\columnwidth]{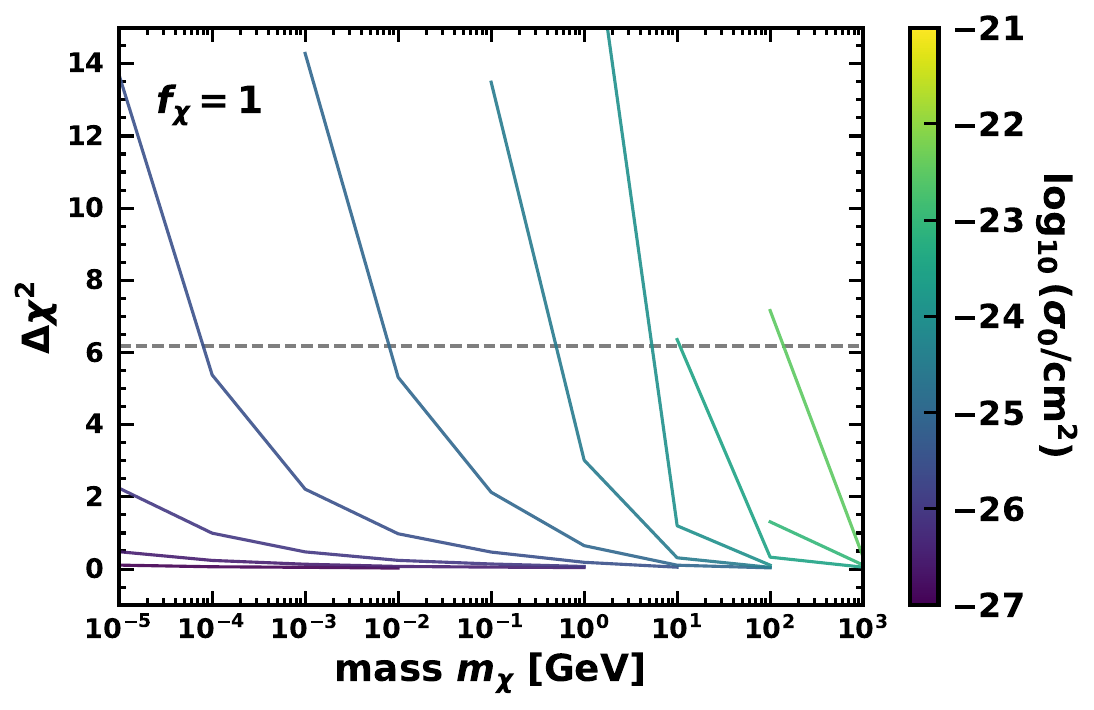}
    \caption{ 
    Profile likelihoods of the mass at fixed cross sections, (indicated by the color bar), interpolated from the profile likelihoods of the cross section at fixed mass, for fixed interaction fraction $f_\chi=1$. The profile likelihood for the mass flattens as the cross section goes to zero.  
    }
    \label{fig:massprofiles}
\end{figure}

We show in Fig.~\ref{fig:massprofiles} the mass profile likelihoods, plotted by interpolating the cross section profile likelihoods for fixed mass to instead show mass profile likelihoods for fixed cross section. 
At smaller cross sections, the likelihood is flat in mass since the mass barely affects the fit and moving in the mass direction does not degrade the likelihood. At larger cross sections, the mass profile is one-sided rather than symmetric, showing a one-sided information constraint on the mass (not a physical boundary as with the cross section), ultimately departing from a clean parabolic form. This shows that mass does not contribute a clean additional degree of freedom to the $\chi^2$ distribution of the log-likelihood-ratio statistic in this regime. Only where the mass is locally parabolic and not at small cross sections will it contribute a clean additional degree of freedom to the $\chi^2$ distribution of the log-likelihood ratio statistic.

Simulating mock CMB observations or precisely quantifying the appropriate mixture of $\chi^2$ distributions for our two free parameters is beyond the scope of this work. 
Since the true cut lies between 4 (were mass to contribute no clean degree of freedom) and 6.18 (two Gaussian degrees of freedom), and because a larger cut yields a looser upper limit, 6.18 is the most conservative choice; any more appropriate cut would only tighten our limits. Additionally, the boundary correction could lead to tighter constraints, but we find that the boundary correction to the 1D per-mass constraint on the cross section makes only a minor difference. 
Ultimately, we choose to present 2D constraints using the naive graphical method for two free parameters, when $\Delta\chi^2=6.18$. 

\begin{figure}
    \centering
    \includegraphics[width=\columnwidth]{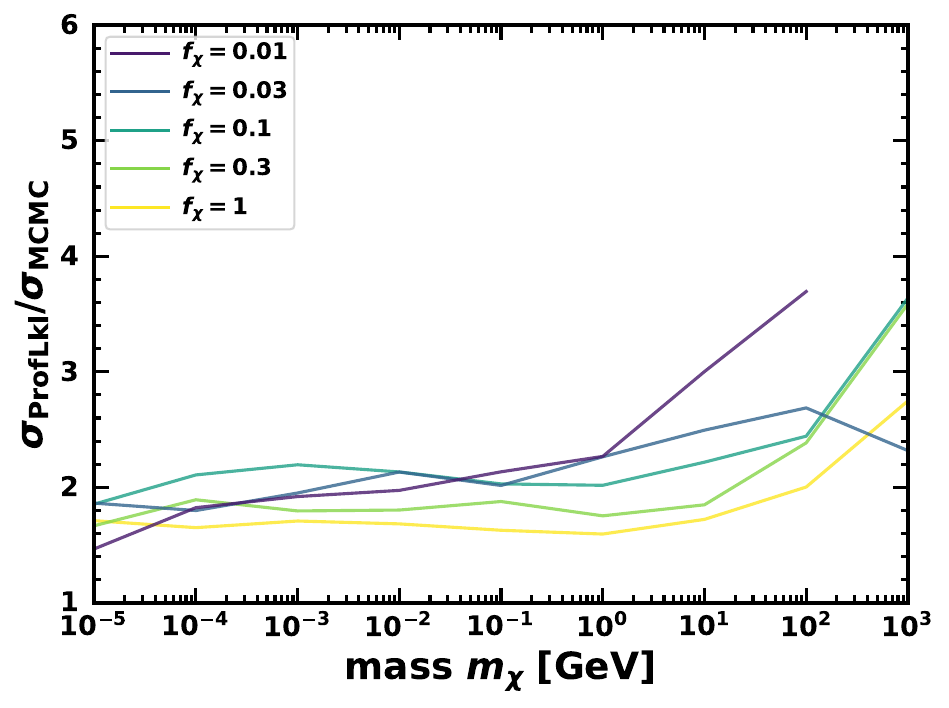}
    \caption{The ratio of the profile-likelihood 1D per-mass constraints on the cross section using the boundary-corrected graphical method, to the Bayesian MCMC constraints. 
    In the case where the likelihood is flat in the mass profile, the mass as a free parameter does not contribute an additional degree of freedom to the $\chi^2$ distribution of the log-likelihood ratio statistic, in which case the appropriate $\Delta\chi^2$ cut would match that of the 1D per-mass case. 
    The likelihood is not flat in the mass profile (see Fig.~\ref{fig:massprofiles}), but we present these results as the tightest-possible appropriate profile-likelihood constraints, which are still not as constraining as the Bayesian constraints.
    }
    \label{fig:FreqBayesRatio_masked_1D_FC_Herold}
\end{figure}

To test whether this choice impacts our results, we take the 1D per-mass boundary-corrected limit as an absolute floor---tighter than any defensible 2D profile-likelihood constraint---and show that even this constraint is not as tight as the Bayesian MCMC constraints. Figure~\ref{fig:FreqBayesRatio_masked_1D_FC_Herold} shows the comparison of our most conservative limiting case, the 1D per-mass boundary-corrected limit (which bounds any 2D limit) 
to the Bayesian constraints.
We emphasize that we do not expect the distribution to be near that of the 1D per-mass boundary-corrected case. 
We present this only to show that even neglecting the influence of the free mass on the distribution of the log-likelihood ratio statistic does not result in constraints that are as tight as those obtained from the Bayesian method. Therefore, our conclusions regarding the comparison of these two statistical methods does not depend on the details of the confidence region construction. The Bayesian constraints are tighter than the profile-likelihood constraints regardless of the precise $\Delta\chi^2$ cut and the details of the boundary correction.

\section{Gaussian fits to profile likelihoods}\label{app:GaussProfLkl}

\begin{figure*}
    \centering
    \includegraphics[width=\columnwidth]{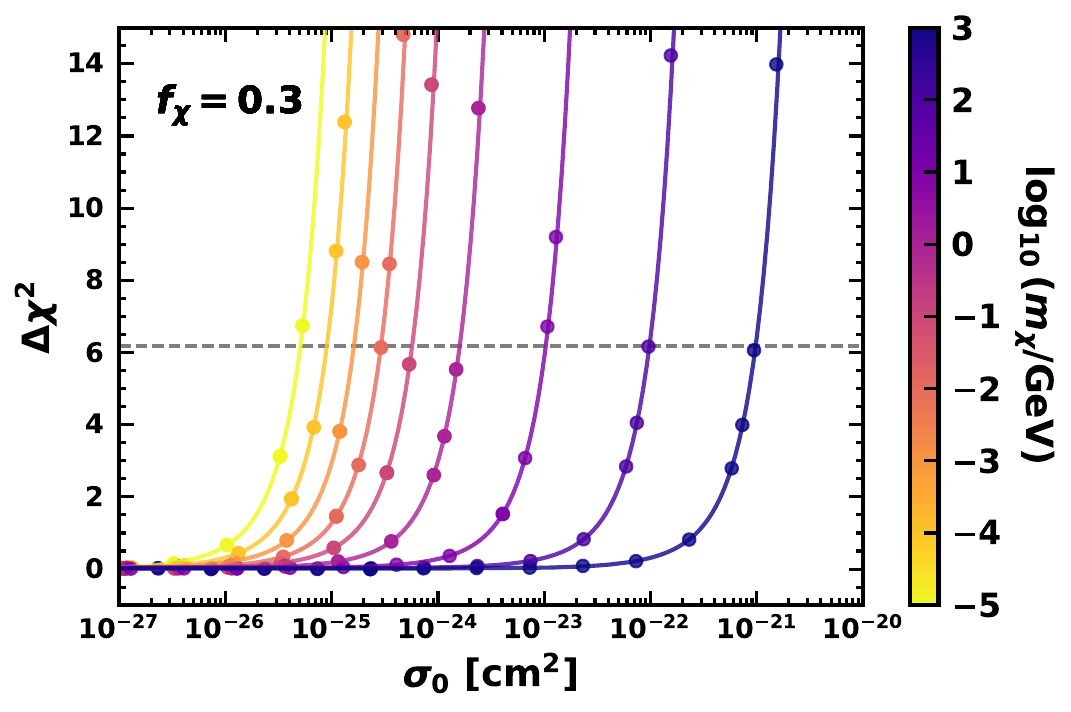}
    \includegraphics[width=\columnwidth]{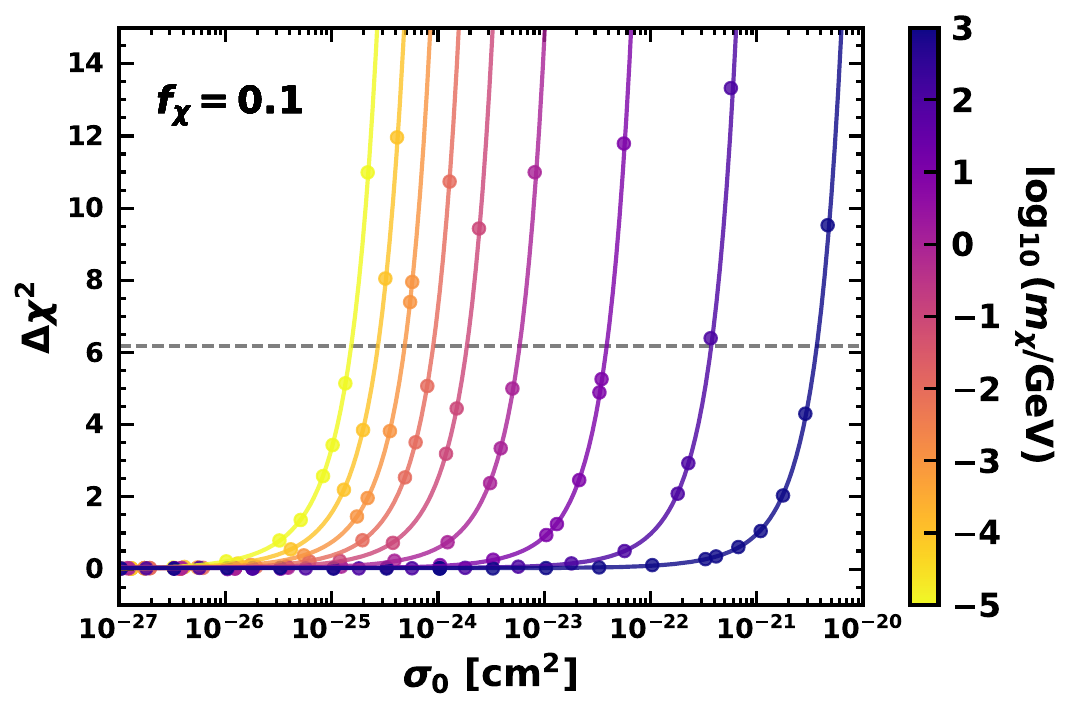}    \includegraphics[width=\columnwidth]{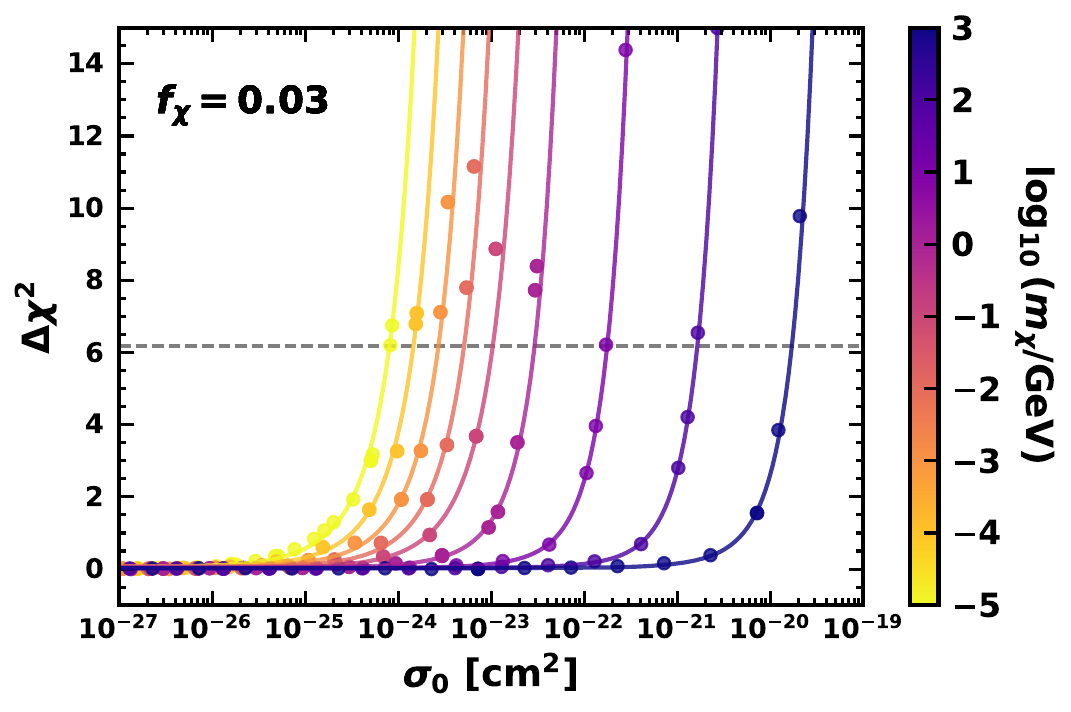}    \includegraphics[width=\columnwidth]{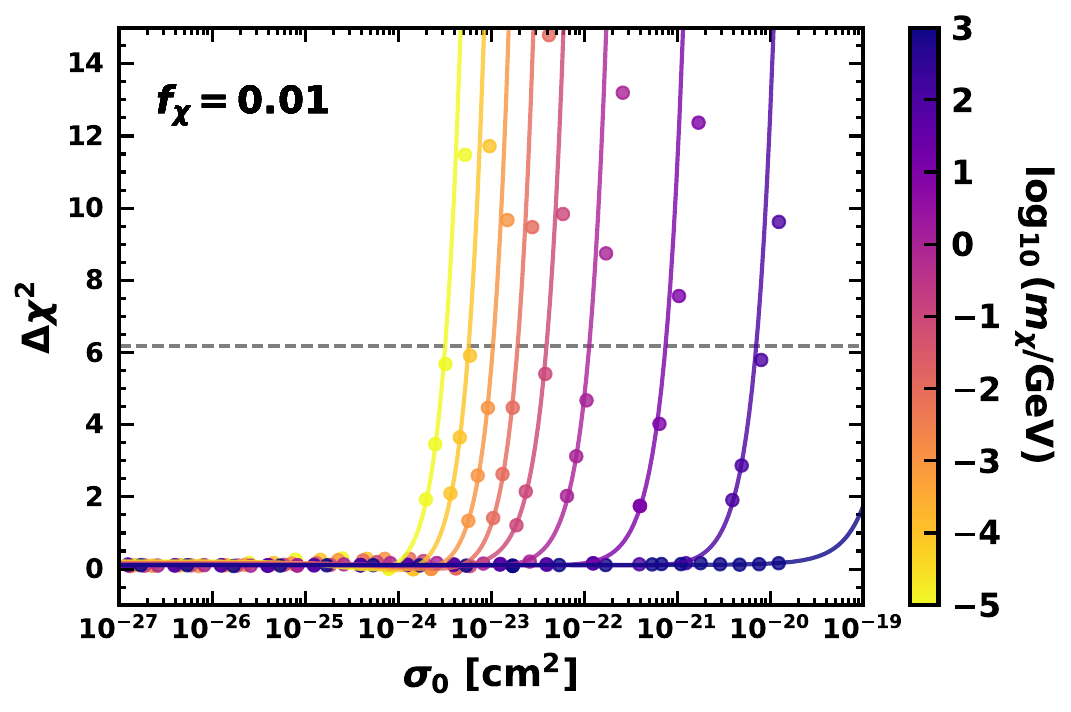}
    \caption{Profile likelihoods of the cross section $\sigma_0$ for different fixed dark matter particle masses $m_\chi$, as indicated by the color bar, and fixed interaction fraction $f_\chi$, as labeled on the individual plots. Points indicate likelihood values, while the lines show the parabolic fit to $\Delta\chi^2$. 
    The horizontal line shows the $\Delta\chi^2=6.18$ cutoff value for a 95.45\% C.L., illustrating where the upper limits on the cross section are drawn using the graphical method for constructing confidence regions.
    }
    \label{fig:ProfileLikelihoods_f}
\end{figure*}

In Fig.~\ref{fig:ProfileLikelihoods}, we show the parabolic fits to the profile likelihood points for the $f_\chi=1$ case. For completeness, we provide the analogous plots for the fractional cases with $f_\chi < 1$ in Fig.~\ref{fig:ProfileLikelihoods_f}. The points indicate the likelihoods computed with \texttt{Procoli}. 
The lines show the parabolic fit to $\Delta\chi^2(\sigma_0)$, obtained by taking the profile likelihood in the form $e^{-\Delta\chi^2(\sigma_0)/2}$ and fitting a Gaussian for linearly-spaced $\sigma_0$ values. 
We fit the Gaussian to points where $\Delta\chi^2<10$, since the profile likelihood is only expected to be Gaussian near the MLE. 
The Gaussian extrapolates to negative values of the cross section. We use the graphical method to set limits on the cross section based on where the parabolic fit intersects the cutoff value $\Delta\chi^2=6.18$, which obtains a 95.45\% confidence region given our model with two parameters of interest.

%%%%%%%%%%%%%%%%%%%%%%%%%%%%%%%%%%%%%%%%%%%%%%%%%%%%%%%%%%%%%%%%%%%%%%%%%%%%%%%
\bibliography{bib}

\begin{thebibliography}{76}%
\makeatletter
\providecommand \@ifxundefined [1]{%
 \@ifx{#1\undefined}
}%
\providecommand \@ifnum [1]{%
 \ifnum #1\expandafter \@firstoftwo
 \else \expandafter \@secondoftwo
 \fi
}%
\providecommand \@ifx [1]{%
 \ifx #1\expandafter \@firstoftwo
 \else \expandafter \@secondoftwo
 \fi
}%
\providecommand \natexlab [1]{#1}%
\providecommand \enquote  [1]{``#1''}%
\providecommand \bibnamefont  [1]{#1}%
\providecommand \bibfnamefont [1]{#1}%
\providecommand \citenamefont [1]{#1}%
\providecommand \href@noop [0]{\@secondoftwo}%
\providecommand \href [0]{\begingroup \@sanitize@url \@href}%
\providecommand \@href[1]{\@@startlink{#1}\@@href}%
\providecommand \@@href[1]{\endgroup#1\@@endlink}%
\providecommand \@sanitize@url [0]{\catcode `\\12\catcode `\$12\catcode `\&12\catcode `\#12\catcode `\^12\catcode `\_12\catcode `\%12\relax}%
\providecommand \@@startlink[1]{}%
\providecommand \@@endlink[0]{}%
\providecommand \url  [0]{\begingroup\@sanitize@url \@url }%
\providecommand \@url [1]{\endgroup\@href {#1}{\urlprefix }}%
\providecommand \urlprefix  [0]{URL }%
\providecommand \Eprint [0]{\href }%
\providecommand \doibase [0]{https://doi.org/}%
\providecommand \selectlanguage [0]{\@gobble}%
\providecommand \bibinfo  [0]{\@secondoftwo}%
\providecommand \bibfield  [0]{\@secondoftwo}%
\providecommand \translation [1]{[#1]}%
\providecommand \BibitemOpen [0]{}%
\providecommand \bibitemStop [0]{}%
\providecommand \bibitemNoStop [0]{.\EOS\space}%
\providecommand \EOS [0]{\spacefactor3000\relax}%
\providecommand \BibitemShut  [1]{\csname bibitem#1\endcsname}%
\let\auto@bib@innerbib\@empty
%</preamble>
\bibitem [{\citenamefont {Aghanim}\ \emph {et~al.}(2020)\citenamefont {Aghanim} \emph {et~al.}}]{PlanckCollaboration_2020_Planck}%
  \BibitemOpen
  \bibfield  {author} {\bibinfo {author} {\bibfnamefont {N.}~\bibnamefont {Aghanim}} \emph {et~al.} (\bibinfo {collaboration} {Planck}),\ }\bibfield  {title} {\bibinfo {title} {{Planck 2018 results. V. CMB power spectra and likelihoods}},\ }\href {https://doi.org/10.1051/0004-6361/201936386} {\bibfield  {journal} {\bibinfo  {journal} {Astron. Astrophys.}\ }\textbf {\bibinfo {volume} {641}},\ \bibinfo {pages} {A5} (\bibinfo {year} {2020})},\ \Eprint {https://arxiv.org/abs/1907.12875} {arXiv:1907.12875 [astro-ph.CO]} \BibitemShut {NoStop}%
\bibitem [{\citenamefont {Miyatake}\ \emph {et~al.}(2023)\citenamefont {Miyatake} \emph {et~al.}}]{Miyatake:2023njf}%
  \BibitemOpen
  \bibfield  {author} {\bibinfo {author} {\bibfnamefont {H.}~\bibnamefont {Miyatake}} \emph {et~al.},\ }\bibfield  {title} {\bibinfo {title} {{Hyper Suprime-Cam Year 3 results: Cosmology from galaxy clustering and weak lensing with HSC and SDSS using the emulator based halo model}},\ }\href {https://doi.org/10.1103/PhysRevD.108.123517} {\bibfield  {journal} {\bibinfo  {journal} {Phys. Rev. D}\ }\textbf {\bibinfo {volume} {108}},\ \bibinfo {pages} {123517} (\bibinfo {year} {2023})},\ \Eprint {https://arxiv.org/abs/2304.00704} {arXiv:2304.00704 [astro-ph.CO]} \BibitemShut {NoStop}%
\bibitem [{\citenamefont {Bocquet}\ \emph {et~al.}(2024)\citenamefont {Bocquet} \emph {et~al.}}]{SPT_2024}%
  \BibitemOpen
  \bibfield  {author} {\bibinfo {author} {\bibfnamefont {S.}~\bibnamefont {Bocquet}} \emph {et~al.} (\bibinfo {collaboration} {SPT, DES}),\ }\bibfield  {title} {\bibinfo {title} {{SPT clusters with DES and HST weak lensing. II. Cosmological constraints from the abundance of massive halos}},\ }\href {https://doi.org/10.1103/PhysRevD.110.083510} {\bibfield  {journal} {\bibinfo  {journal} {Phys. Rev. D}\ }\textbf {\bibinfo {volume} {110}},\ \bibinfo {pages} {083510} (\bibinfo {year} {2024})},\ \Eprint {https://arxiv.org/abs/2401.02075} {arXiv:2401.02075 [astro-ph.CO]} \BibitemShut {NoStop}%
\bibitem [{\citenamefont {Calabrese}\ \emph {et~al.}(2025)\citenamefont {Calabrese} \emph {et~al.}}]{AtacamaCosmologyTelescope_2025}%
  \BibitemOpen
  \bibfield  {author} {\bibinfo {author} {\bibfnamefont {E.}~\bibnamefont {Calabrese}} \emph {et~al.} (\bibinfo {collaboration} {Atacama Cosmology Telescope}),\ }\bibfield  {title} {\bibinfo {title} {{The Atacama Cosmology Telescope: DR6 constraints on extended cosmological models}},\ }\href {https://doi.org/10.1088/1475-7516/2025/11/063} {\bibfield  {journal} {\bibinfo  {journal} {JCAP}\ }\textbf {\bibinfo {volume} {11}},\ \bibinfo {pages} {063}},\ \Eprint {https://arxiv.org/abs/2503.14454} {arXiv:2503.14454 [astro-ph.CO]} \BibitemShut {NoStop}%
\bibitem [{\citenamefont {Wright}\ \emph {et~al.}(2025)\citenamefont {Wright} \emph {et~al.}}]{Wright:2025xka}%
  \BibitemOpen
  \bibfield  {author} {\bibinfo {author} {\bibfnamefont {A.~H.}\ \bibnamefont {Wright}} \emph {et~al.},\ }\bibfield  {title} {\bibinfo {title} {{KiDS-Legacy: Cosmological constraints from cosmic shear with the complete Kilo-Degree Survey}},\ }\href {https://doi.org/10.1051/0004-6361/202554908} {\bibfield  {journal} {\bibinfo  {journal} {Astron. Astrophys.}\ }\textbf {\bibinfo {volume} {703}},\ \bibinfo {pages} {A158} (\bibinfo {year} {2025})},\ \Eprint {https://arxiv.org/abs/2503.19441} {arXiv:2503.19441 [astro-ph.CO]} \BibitemShut {NoStop}%
\bibitem [{\citenamefont {Abbott}\ \emph {et~al.}(2026{\natexlab{a}})\citenamefont {Abbott} \emph {et~al.}}]{DES_2026}%
  \BibitemOpen
  \bibfield  {author} {\bibinfo {author} {\bibfnamefont {T.~M.~C.}\ \bibnamefont {Abbott}} \emph {et~al.} (\bibinfo {collaboration} {DES}),\ }\bibfield  {title} {\bibinfo {title} {{Dark Energy Survey Year 6 Results: Cosmological Constraints from Galaxy Clustering and Weak Lensing}},\ }\href@noop {} {\  (\bibinfo {year} {2026}{\natexlab{a}})},\ \Eprint {https://arxiv.org/abs/2601.14559} {arXiv:2601.14559 [astro-ph.CO]} \BibitemShut {NoStop}%
\bibitem [{\citenamefont {Abbott}\ \emph {et~al.}(2026{\natexlab{b}})\citenamefont {Abbott} \emph {et~al.}}]{DES:2026mkc}%
  \BibitemOpen
  \bibfield  {author} {\bibinfo {author} {\bibfnamefont {T.~M.~C.}\ \bibnamefont {Abbott}} \emph {et~al.} (\bibinfo {collaboration} {DES}),\ }\bibfield  {title} {\bibinfo {title} {{Dark Energy Survey Year 6 Results: Cosmological Constraints from Cosmic Shear}},\ }\href@noop {} {\  (\bibinfo {year} {2026}{\natexlab{b}})},\ \Eprint {https://arxiv.org/abs/2602.10065} {arXiv:2602.10065 [astro-ph.CO]} \BibitemShut {NoStop}%
\bibitem [{\citenamefont {{Cushman}}\ \emph {et~al.}(2013)\citenamefont {{Cushman}}, \citenamefont {{Galbiati}}, \citenamefont {{McKinsey}}, \citenamefont {{Robertson}}, \citenamefont {{Tait}}, \citenamefont {{Bauer}}, \citenamefont {{Borgland}}, \citenamefont {{Cabrera}}, \citenamefont {{Calaprice}}, \citenamefont {{Cooley}}, \citenamefont {{Empl}}, \citenamefont {{Essig}}, \citenamefont {{Figueroa-Feliciano}}, \citenamefont {{Gaitskell}}, \citenamefont {{Golwala}}, \citenamefont {{Hall}}, \citenamefont {{Hill}}, \citenamefont {{Hime}}, \citenamefont {{Hoppe}}, \citenamefont {{Hsu}}, \citenamefont {{Hungerford}}, \citenamefont {{Jacobsen}}, \citenamefont {{Kelsey}}, \citenamefont {{Lang}}, \citenamefont {{Lippincott}}, \citenamefont {{Loer}}, \citenamefont {{Luitz}}, \citenamefont {{Mandic}}, \citenamefont {{Mardon}}, \citenamefont {{Maricic}}, \citenamefont {{Maruyama}}, \citenamefont {{Mahapatra}}, \citenamefont {{Nelson}}, \citenamefont {{Orrell}}, \citenamefont {{Palladino}}, \citenamefont {{Pantic}},
  \citenamefont {{Partridge}}, \citenamefont {{Ryd}}, \citenamefont {{Saab}}, \citenamefont {{Sadoulet}}, \citenamefont {{Schnee}}, \citenamefont {{Shepherd}}, \citenamefont {{Sonnenschein}}, \citenamefont {{Sorensen}}, \citenamefont {{Szydagis}}, \citenamefont {{Volansky}}, \citenamefont {{Witherell}}, \citenamefont {{Wright}},\ and\ \citenamefont {{Zurek}}}]{Cushman_2013_snowmass}%
  \BibitemOpen
  \bibfield  {author} {\bibinfo {author} {\bibfnamefont {P.}~\bibnamefont {{Cushman}}}, \bibinfo {author} {\bibfnamefont {C.}~\bibnamefont {{Galbiati}}}, \bibinfo {author} {\bibfnamefont {D.~N.}\ \bibnamefont {{McKinsey}}}, \bibinfo {author} {\bibfnamefont {H.}~\bibnamefont {{Robertson}}}, \bibinfo {author} {\bibfnamefont {T.~M.~P.}\ \bibnamefont {{Tait}}}, \bibinfo {author} {\bibfnamefont {D.}~\bibnamefont {{Bauer}}}, \bibinfo {author} {\bibfnamefont {A.}~\bibnamefont {{Borgland}}}, \bibinfo {author} {\bibfnamefont {B.}~\bibnamefont {{Cabrera}}}, \bibinfo {author} {\bibfnamefont {F.}~\bibnamefont {{Calaprice}}}, \bibinfo {author} {\bibfnamefont {J.}~\bibnamefont {{Cooley}}}, \bibinfo {author} {\bibfnamefont {T.}~\bibnamefont {{Empl}}}, \bibinfo {author} {\bibfnamefont {R.}~\bibnamefont {{Essig}}}, \bibinfo {author} {\bibfnamefont {E.}~\bibnamefont {{Figueroa-Feliciano}}}, \bibinfo {author} {\bibfnamefont {R.}~\bibnamefont {{Gaitskell}}}, \bibinfo {author} {\bibfnamefont {S.}~\bibnamefont {{Golwala}}},
  \bibinfo {author} {\bibfnamefont {J.}~\bibnamefont {{Hall}}}, \bibinfo {author} {\bibfnamefont {R.}~\bibnamefont {{Hill}}}, \bibinfo {author} {\bibfnamefont {A.}~\bibnamefont {{Hime}}}, \bibinfo {author} {\bibfnamefont {E.}~\bibnamefont {{Hoppe}}}, \bibinfo {author} {\bibfnamefont {L.}~\bibnamefont {{Hsu}}}, \bibinfo {author} {\bibfnamefont {E.}~\bibnamefont {{Hungerford}}}, \bibinfo {author} {\bibfnamefont {R.}~\bibnamefont {{Jacobsen}}}, \bibinfo {author} {\bibfnamefont {M.}~\bibnamefont {{Kelsey}}}, \bibinfo {author} {\bibfnamefont {R.~F.}\ \bibnamefont {{Lang}}}, \bibinfo {author} {\bibfnamefont {W.~H.}\ \bibnamefont {{Lippincott}}}, \bibinfo {author} {\bibfnamefont {B.}~\bibnamefont {{Loer}}}, \bibinfo {author} {\bibfnamefont {S.}~\bibnamefont {{Luitz}}}, \bibinfo {author} {\bibfnamefont {V.}~\bibnamefont {{Mandic}}}, \bibinfo {author} {\bibfnamefont {J.}~\bibnamefont {{Mardon}}}, \bibinfo {author} {\bibfnamefont {J.}~\bibnamefont {{Maricic}}}, \bibinfo {author} {\bibfnamefont {R.}~\bibnamefont
  {{Maruyama}}}, \bibinfo {author} {\bibfnamefont {R.}~\bibnamefont {{Mahapatra}}}, \bibinfo {author} {\bibfnamefont {H.}~\bibnamefont {{Nelson}}}, \bibinfo {author} {\bibfnamefont {J.}~\bibnamefont {{Orrell}}}, \bibinfo {author} {\bibfnamefont {K.}~\bibnamefont {{Palladino}}}, \bibinfo {author} {\bibfnamefont {E.}~\bibnamefont {{Pantic}}}, \bibinfo {author} {\bibfnamefont {R.}~\bibnamefont {{Partridge}}}, \bibinfo {author} {\bibfnamefont {A.}~\bibnamefont {{Ryd}}}, \bibinfo {author} {\bibfnamefont {T.}~\bibnamefont {{Saab}}}, \bibinfo {author} {\bibfnamefont {B.}~\bibnamefont {{Sadoulet}}}, \bibinfo {author} {\bibfnamefont {R.}~\bibnamefont {{Schnee}}}, \bibinfo {author} {\bibfnamefont {W.}~\bibnamefont {{Shepherd}}}, \bibinfo {author} {\bibfnamefont {A.}~\bibnamefont {{Sonnenschein}}}, \bibinfo {author} {\bibfnamefont {P.}~\bibnamefont {{Sorensen}}}, \bibinfo {author} {\bibfnamefont {M.}~\bibnamefont {{Szydagis}}}, \bibinfo {author} {\bibfnamefont {T.}~\bibnamefont {{Volansky}}}, \bibinfo {author}
  {\bibfnamefont {M.}~\bibnamefont {{Witherell}}}, \bibinfo {author} {\bibfnamefont {D.}~\bibnamefont {{Wright}}},\ and\ \bibinfo {author} {\bibfnamefont {K.}~\bibnamefont {{Zurek}}},\ }\bibfield  {title} {\bibinfo {title} {{Snowmass CF1 Summary: WIMP Dark Matter Direct Detection}},\ }\href {https://doi.org/10.48550/arXiv.1310.8327} {\bibfield  {journal} {\bibinfo  {journal} {arXiv e-prints}\ ,\ \bibinfo {eid} {arXiv:1310.8327}} (\bibinfo {year} {2013})},\ \Eprint {https://arxiv.org/abs/1310.8327} {arXiv:1310.8327 [hep-ex]} \BibitemShut {NoStop}%
\bibitem [{\citenamefont {Aprile}\ \emph {et~al.}(2023)\citenamefont {Aprile} \emph {et~al.}}]{XENON_2023_First}%
  \BibitemOpen
  \bibfield  {author} {\bibinfo {author} {\bibfnamefont {E.}~\bibnamefont {Aprile}} \emph {et~al.} (\bibinfo {collaboration} {XENON}),\ }\bibfield  {title} {\bibinfo {title} {{First Dark Matter Search with Nuclear Recoils from the XENONnT Experiment}},\ }\href {https://doi.org/10.1103/PhysRevLett.131.041003} {\bibfield  {journal} {\bibinfo  {journal} {Phys. Rev. Lett.}\ }\textbf {\bibinfo {volume} {131}},\ \bibinfo {pages} {041003} (\bibinfo {year} {2023})},\ \Eprint {https://arxiv.org/abs/2303.14729} {arXiv:2303.14729 [hep-ex]} \BibitemShut {NoStop}%
\bibitem [{\citenamefont {Bo}\ \emph {et~al.}(2025)\citenamefont {Bo} \emph {et~al.}}]{PandaX_2024_Dark}%
  \BibitemOpen
  \bibfield  {author} {\bibinfo {author} {\bibfnamefont {Z.}~\bibnamefont {Bo}} \emph {et~al.} (\bibinfo {collaboration} {PandaX}),\ }\bibfield  {title} {\bibinfo {title} {{Dark Matter Search Results from 1.54{\,}{\,}Tonne{\textperiodcentered}Year Exposure of PandaX-4T}},\ }\href {https://doi.org/10.1103/PhysRevLett.134.011805} {\bibfield  {journal} {\bibinfo  {journal} {Phys. Rev. Lett.}\ }\textbf {\bibinfo {volume} {134}},\ \bibinfo {pages} {011805} (\bibinfo {year} {2025})},\ \Eprint {https://arxiv.org/abs/2408.00664} {arXiv:2408.00664 [hep-ex]} \BibitemShut {NoStop}%
\bibitem [{\citenamefont {Aalbers}\ \emph {et~al.}(2025)\citenamefont {Aalbers} \emph {et~al.}}]{LZ_2024_Dark}%
  \BibitemOpen
  \bibfield  {author} {\bibinfo {author} {\bibfnamefont {J.}~\bibnamefont {Aalbers}} \emph {et~al.} (\bibinfo {collaboration} {LZ}),\ }\bibfield  {title} {\bibinfo {title} {{Dark Matter Search Results from 4.2{\,}{\,}Tonne-Years of Exposure of the LUX-ZEPLIN (LZ) Experiment}},\ }\href {https://doi.org/10.1103/4dyc-z8zf} {\bibfield  {journal} {\bibinfo  {journal} {Phys. Rev. Lett.}\ }\textbf {\bibinfo {volume} {135}},\ \bibinfo {pages} {011802} (\bibinfo {year} {2025})},\ \Eprint {https://arxiv.org/abs/2410.17036} {arXiv:2410.17036 [hep-ex]} \BibitemShut {NoStop}%
\bibitem [{\citenamefont {Emken}\ and\ \citenamefont {Kouvaris}(2017)}]{Emken_2017_DaMaSCUS}%
  \BibitemOpen
  \bibfield  {author} {\bibinfo {author} {\bibfnamefont {T.}~\bibnamefont {Emken}}\ and\ \bibinfo {author} {\bibfnamefont {C.}~\bibnamefont {Kouvaris}},\ }\bibfield  {title} {\bibinfo {title} {{DaMaSCUS: The Impact of Underground Scatterings on Direct Detection of Light Dark Matter}},\ }\href {https://doi.org/10.1088/1475-7516/2017/10/031} {\bibfield  {journal} {\bibinfo  {journal} {JCAP}\ }\textbf {\bibinfo {volume} {10}},\ \bibinfo {pages} {031}},\ \Eprint {https://arxiv.org/abs/1706.02249} {arXiv:1706.02249 [hep-ph]} \BibitemShut {NoStop}%
\bibitem [{\citenamefont {Boddy}\ and\ \citenamefont {Gluscevic}(2018)}]{BoddyGluscevic_2018_first}%
  \BibitemOpen
  \bibfield  {author} {\bibinfo {author} {\bibfnamefont {K.~K.}\ \bibnamefont {Boddy}}\ and\ \bibinfo {author} {\bibfnamefont {V.}~\bibnamefont {Gluscevic}},\ }\bibfield  {title} {\bibinfo {title} {{First Cosmological Constraint on the Effective Theory of Dark Matter-Proton Interactions}},\ }\href {https://doi.org/10.1103/PhysRevD.98.083510} {\bibfield  {journal} {\bibinfo  {journal} {Phys. Rev. D}\ }\textbf {\bibinfo {volume} {98}},\ \bibinfo {pages} {083510} (\bibinfo {year} {2018})},\ \Eprint {https://arxiv.org/abs/1801.08609} {arXiv:1801.08609 [astro-ph.CO]} \BibitemShut {NoStop}%
\bibitem [{\citenamefont {Chen}\ \emph {et~al.}(2002)\citenamefont {Chen}, \citenamefont {Hannestad},\ and\ \citenamefont {Scherrer}}]{Chen_2002_cosmic}%
  \BibitemOpen
  \bibfield  {author} {\bibinfo {author} {\bibfnamefont {X.-l.}\ \bibnamefont {Chen}}, \bibinfo {author} {\bibfnamefont {S.}~\bibnamefont {Hannestad}},\ and\ \bibinfo {author} {\bibfnamefont {R.~J.}\ \bibnamefont {Scherrer}},\ }\bibfield  {title} {\bibinfo {title} {{Cosmic microwave background and large scale structure limits on the interaction between dark matter and baryons}},\ }\href {https://doi.org/10.1103/PhysRevD.65.123515} {\bibfield  {journal} {\bibinfo  {journal} {Phys. Rev. D}\ }\textbf {\bibinfo {volume} {65}},\ \bibinfo {pages} {123515} (\bibinfo {year} {2002})},\ \Eprint {https://arxiv.org/abs/astro-ph/0202496} {arXiv:astro-ph/0202496} \BibitemShut {NoStop}%
\bibitem [{\citenamefont {Dvorkin}\ \emph {et~al.}(2014)\citenamefont {Dvorkin}, \citenamefont {Blum},\ and\ \citenamefont {Kamionkowski}}]{Dvorkin_2013_constraining}%
  \BibitemOpen
  \bibfield  {author} {\bibinfo {author} {\bibfnamefont {C.}~\bibnamefont {Dvorkin}}, \bibinfo {author} {\bibfnamefont {K.}~\bibnamefont {Blum}},\ and\ \bibinfo {author} {\bibfnamefont {M.}~\bibnamefont {Kamionkowski}},\ }\bibfield  {title} {\bibinfo {title} {{Constraining Dark Matter-Baryon Scattering with Linear Cosmology}},\ }\href {https://doi.org/10.1103/PhysRevD.89.023519} {\bibfield  {journal} {\bibinfo  {journal} {Phys. Rev. D}\ }\textbf {\bibinfo {volume} {89}},\ \bibinfo {pages} {023519} (\bibinfo {year} {2014})},\ \Eprint {https://arxiv.org/abs/1311.2937} {arXiv:1311.2937 [astro-ph.CO]} \BibitemShut {NoStop}%
\bibitem [{\citenamefont {Boddy}\ \emph {et~al.}(2018)\citenamefont {Boddy}, \citenamefont {Gluscevic}, \citenamefont {Poulin}, \citenamefont {Kovetz}, \citenamefont {Kamionkowski},\ and\ \citenamefont {Barkana}}]{Boddy_2018_new}%
  \BibitemOpen
  \bibfield  {author} {\bibinfo {author} {\bibfnamefont {K.~K.}\ \bibnamefont {Boddy}}, \bibinfo {author} {\bibfnamefont {V.}~\bibnamefont {Gluscevic}}, \bibinfo {author} {\bibfnamefont {V.}~\bibnamefont {Poulin}}, \bibinfo {author} {\bibfnamefont {E.~D.}\ \bibnamefont {Kovetz}}, \bibinfo {author} {\bibfnamefont {M.}~\bibnamefont {Kamionkowski}},\ and\ \bibinfo {author} {\bibfnamefont {R.}~\bibnamefont {Barkana}},\ }\bibfield  {title} {\bibinfo {title} {{Critical assessment of CMB limits on dark matter-baryon scattering: New treatment of the relative bulk velocity}},\ }\href {https://doi.org/10.1103/PhysRevD.98.123506} {\bibfield  {journal} {\bibinfo  {journal} {Phys. Rev. D}\ }\textbf {\bibinfo {volume} {98}},\ \bibinfo {pages} {123506} (\bibinfo {year} {2018})},\ \Eprint {https://arxiv.org/abs/1808.00001} {arXiv:1808.00001 [astro-ph.CO]} \BibitemShut {NoStop}%
\bibitem [{\citenamefont {{Gluscevic}}\ and\ \citenamefont {{Boddy}}(2018)}]{GluscevicBoddy_2018_constraints}%
  \BibitemOpen
  \bibfield  {author} {\bibinfo {author} {\bibfnamefont {V.}~\bibnamefont {{Gluscevic}}}\ and\ \bibinfo {author} {\bibfnamefont {K.~K.}\ \bibnamefont {{Boddy}}},\ }\bibfield  {title} {\bibinfo {title} {{Constraints on Scattering of keV-TeV Dark Matter with Protons in the Early Universe}},\ }\href {https://doi.org/10.1103/PhysRevLett.121.081301} {\bibfield  {journal} {\bibinfo  {journal} {\prl}\ }\textbf {\bibinfo {volume} {121}},\ \bibinfo {eid} {081301} (\bibinfo {year} {2018})},\ \Eprint {https://arxiv.org/abs/1712.07133} {arXiv:1712.07133 [astro-ph.CO]} \BibitemShut {NoStop}%
\bibitem [{\citenamefont {Xu}\ \emph {et~al.}(2018)\citenamefont {Xu}, \citenamefont {Dvorkin},\ and\ \citenamefont {Chael}}]{Xu_2018_Probing}%
  \BibitemOpen
  \bibfield  {author} {\bibinfo {author} {\bibfnamefont {W.~L.}\ \bibnamefont {Xu}}, \bibinfo {author} {\bibfnamefont {C.}~\bibnamefont {Dvorkin}},\ and\ \bibinfo {author} {\bibfnamefont {A.}~\bibnamefont {Chael}},\ }\bibfield  {title} {\bibinfo {title} {{Probing sub-GeV Dark Matter-Baryon Scattering with Cosmological Observables}},\ }\href {https://doi.org/10.1103/PhysRevD.97.103530} {\bibfield  {journal} {\bibinfo  {journal} {Phys. Rev. D}\ }\textbf {\bibinfo {volume} {97}},\ \bibinfo {pages} {103530} (\bibinfo {year} {2018})},\ \Eprint {https://arxiv.org/abs/1802.06788} {arXiv:1802.06788 [astro-ph.CO]} \BibitemShut {NoStop}%
\bibitem [{\citenamefont {{Nguyen}}\ \emph {et~al.}(2021)\citenamefont {{Nguyen}}, \citenamefont {{Sarnaaik}}, \citenamefont {{Boddy}}, \citenamefont {{Nadler}},\ and\ \citenamefont {{Gluscevic}}}]{Nguyen_2021_observational}%
  \BibitemOpen
  \bibfield  {author} {\bibinfo {author} {\bibfnamefont {D.~V.}\ \bibnamefont {{Nguyen}}}, \bibinfo {author} {\bibfnamefont {D.}~\bibnamefont {{Sarnaaik}}}, \bibinfo {author} {\bibfnamefont {K.~K.}\ \bibnamefont {{Boddy}}}, \bibinfo {author} {\bibfnamefont {E.~O.}\ \bibnamefont {{Nadler}}},\ and\ \bibinfo {author} {\bibfnamefont {V.}~\bibnamefont {{Gluscevic}}},\ }\bibfield  {title} {\bibinfo {title} {{Observational constraints on dark matter scattering with electrons}},\ }\href {https://doi.org/10.1103/PhysRevD.104.103521} {\bibfield  {journal} {\bibinfo  {journal} {\prd}\ }\textbf {\bibinfo {volume} {104}},\ \bibinfo {eid} {103521} (\bibinfo {year} {2021})},\ \Eprint {https://arxiv.org/abs/2107.12380} {arXiv:2107.12380 [astro-ph.CO]} \BibitemShut {NoStop}%
\bibitem [{\citenamefont {Buen-Abad}\ \emph {et~al.}(2022)\citenamefont {Buen-Abad}, \citenamefont {Essig}, \citenamefont {McKeen},\ and\ \citenamefont {Zhong}}]{Buen-Abad_2022_Cosmological}%
  \BibitemOpen
  \bibfield  {author} {\bibinfo {author} {\bibfnamefont {M.~A.}\ \bibnamefont {Buen-Abad}}, \bibinfo {author} {\bibfnamefont {R.}~\bibnamefont {Essig}}, \bibinfo {author} {\bibfnamefont {D.}~\bibnamefont {McKeen}},\ and\ \bibinfo {author} {\bibfnamefont {Y.-M.}\ \bibnamefont {Zhong}},\ }\bibfield  {title} {\bibinfo {title} {{Cosmological constraints on dark matter interactions with ordinary matter}},\ }\href {https://doi.org/10.1016/j.physrep.2022.02.006} {\bibfield  {journal} {\bibinfo  {journal} {Phys. Rept.}\ }\textbf {\bibinfo {volume} {961}},\ \bibinfo {pages} {1} (\bibinfo {year} {2022})},\ \Eprint {https://arxiv.org/abs/2107.12377} {arXiv:2107.12377 [astro-ph.CO]} \BibitemShut {NoStop}%
\bibitem [{\citenamefont {Boddy}\ \emph {et~al.}(2022)\citenamefont {Boddy}, \citenamefont {Krnjaic},\ and\ \citenamefont {Moltner}}]{Boddy_2022_Investigation}%
  \BibitemOpen
  \bibfield  {author} {\bibinfo {author} {\bibfnamefont {K.~K.}\ \bibnamefont {Boddy}}, \bibinfo {author} {\bibfnamefont {G.}~\bibnamefont {Krnjaic}},\ and\ \bibinfo {author} {\bibfnamefont {S.}~\bibnamefont {Moltner}},\ }\bibfield  {title} {\bibinfo {title} {{Investigation of CMB constraints for dark matter-helium scattering}},\ }\href {https://doi.org/10.1103/PhysRevD.106.043510} {\bibfield  {journal} {\bibinfo  {journal} {Phys. Rev. D}\ }\textbf {\bibinfo {volume} {106}},\ \bibinfo {pages} {043510} (\bibinfo {year} {2022})},\ \Eprint {https://arxiv.org/abs/2204.04225} {arXiv:2204.04225 [astro-ph.CO]} \BibitemShut {NoStop}%
\bibitem [{\citenamefont {Li}\ \emph {et~al.}(2023)\citenamefont {Li} \emph {et~al.}}]{Li:2022mdj}%
  \BibitemOpen
  \bibfield  {author} {\bibinfo {author} {\bibfnamefont {Z.}~\bibnamefont {Li}} \emph {et~al.},\ }\bibfield  {title} {\bibinfo {title} {{The Atacama Cosmology Telescope: limits on dark matter-baryon interactions from DR4 power spectra}},\ }\href {https://doi.org/10.1088/1475-7516/2023/02/046} {\bibfield  {journal} {\bibinfo  {journal} {JCAP}\ }\textbf {\bibinfo {volume} {02}},\ \bibinfo {pages} {046}},\ \Eprint {https://arxiv.org/abs/2208.08985} {arXiv:2208.08985 [astro-ph.CO]} \BibitemShut {NoStop}%
\bibitem [{\citenamefont {An}\ \emph {et~al.}(2024)\citenamefont {An}, \citenamefont {Boddy},\ and\ \citenamefont {Gluscevic}}]{An_2024_Interacting}%
  \BibitemOpen
  \bibfield  {author} {\bibinfo {author} {\bibfnamefont {R.}~\bibnamefont {An}}, \bibinfo {author} {\bibfnamefont {K.~K.}\ \bibnamefont {Boddy}},\ and\ \bibinfo {author} {\bibfnamefont {V.}~\bibnamefont {Gluscevic}},\ }\bibfield  {title} {\bibinfo {title} {Interacting light thermal-relic dark matter: Self-consistent cosmological bounds},\ }\href {https://doi.org/10.1103/PhysRevD.109.123522} {\bibfield  {journal} {\bibinfo  {journal} {Phys. Rev. D}\ }\textbf {\bibinfo {volume} {109}},\ \bibinfo {pages} {123522} (\bibinfo {year} {2024})}\BibitemShut {NoStop}%
\bibitem [{\citenamefont {Mu{\~n}oz}\ and\ \citenamefont {Loeb}(2017)}]{Munoz_2017_Constraints}%
  \BibitemOpen
  \bibfield  {author} {\bibinfo {author} {\bibfnamefont {J.~B.}\ \bibnamefont {Mu{\~n}oz}}\ and\ \bibinfo {author} {\bibfnamefont {A.}~\bibnamefont {Loeb}},\ }\bibfield  {title} {\bibinfo {title} {{Constraints on Dark Matter-Baryon Scattering from the Temperature Evolution of the Intergalactic Medium}},\ }\href {https://doi.org/10.1088/1475-7516/2017/11/043} {\bibfield  {journal} {\bibinfo  {journal} {JCAP}\ }\textbf {\bibinfo {volume} {11}},\ \bibinfo {pages} {043}},\ \Eprint {https://arxiv.org/abs/1708.08923} {arXiv:1708.08923 [astro-ph.CO]} \BibitemShut {NoStop}%
\bibitem [{\citenamefont {Ooba}\ \emph {et~al.}(2019)\citenamefont {Ooba}, \citenamefont {Tashiro},\ and\ \citenamefont {Kadota}}]{Ooba_2019_Cosmological}%
  \BibitemOpen
  \bibfield  {author} {\bibinfo {author} {\bibfnamefont {J.}~\bibnamefont {Ooba}}, \bibinfo {author} {\bibfnamefont {H.}~\bibnamefont {Tashiro}},\ and\ \bibinfo {author} {\bibfnamefont {K.}~\bibnamefont {Kadota}},\ }\bibfield  {title} {\bibinfo {title} {{Cosmological constraints on the velocity-dependent baryon-dark matter coupling}},\ }\href {https://doi.org/10.1088/1475-7516/2019/09/020} {\bibfield  {journal} {\bibinfo  {journal} {JCAP}\ }\textbf {\bibinfo {volume} {09}},\ \bibinfo {pages} {020}},\ \Eprint {https://arxiv.org/abs/1902.00826} {arXiv:1902.00826 [astro-ph.CO]} \BibitemShut {NoStop}%
\bibitem [{\citenamefont {Rogers}\ \emph {et~al.}(2022)\citenamefont {Rogers}, \citenamefont {Dvorkin},\ and\ \citenamefont {Peiris}}]{Rogers_2021_Limits}%
  \BibitemOpen
  \bibfield  {author} {\bibinfo {author} {\bibfnamefont {K.~K.}\ \bibnamefont {Rogers}}, \bibinfo {author} {\bibfnamefont {C.}~\bibnamefont {Dvorkin}},\ and\ \bibinfo {author} {\bibfnamefont {H.~V.}\ \bibnamefont {Peiris}},\ }\bibfield  {title} {\bibinfo {title} {{Limits on the Light Dark Matter{\textendash}Proton Cross Section from Cosmic Large-Scale Structure}},\ }\href {https://doi.org/10.1103/PhysRevLett.128.171301} {\bibfield  {journal} {\bibinfo  {journal} {Phys. Rev. Lett.}\ }\textbf {\bibinfo {volume} {128}},\ \bibinfo {pages} {171301} (\bibinfo {year} {2022})},\ \Eprint {https://arxiv.org/abs/2111.10386} {arXiv:2111.10386 [astro-ph.CO]} \BibitemShut {NoStop}%
\bibitem [{\citenamefont {Mu\~noz}\ \emph {et~al.}(2015)\citenamefont {Mu\~noz}, \citenamefont {Kovetz},\ and\ \citenamefont {Ali-Ha\"{\i}moud}}]{Munoz_2015_heating}%
  \BibitemOpen
  \bibfield  {author} {\bibinfo {author} {\bibfnamefont {J.~B.}\ \bibnamefont {Mu\~noz}}, \bibinfo {author} {\bibfnamefont {E.~D.}\ \bibnamefont {Kovetz}},\ and\ \bibinfo {author} {\bibfnamefont {Y.}~\bibnamefont {Ali-Ha\"{\i}moud}},\ }\bibfield  {title} {\bibinfo {title} {Heating of baryons due to scattering with dark matter during the dark ages},\ }\href {https://doi.org/10.1103/PhysRevD.92.083528} {\bibfield  {journal} {\bibinfo  {journal} {Phys. Rev. D}\ }\textbf {\bibinfo {volume} {92}},\ \bibinfo {pages} {083528} (\bibinfo {year} {2015})}\BibitemShut {NoStop}%
\bibitem [{\citenamefont {Slatyer}\ and\ \citenamefont {Wu}(2018)}]{Slatyer_2018_Early}%
  \BibitemOpen
  \bibfield  {author} {\bibinfo {author} {\bibfnamefont {T.~R.}\ \bibnamefont {Slatyer}}\ and\ \bibinfo {author} {\bibfnamefont {C.-L.}\ \bibnamefont {Wu}},\ }\bibfield  {title} {\bibinfo {title} {{Early-Universe constraints on dark matter-baryon scattering and their implications for a global 21 cm signal}},\ }\href {https://doi.org/10.1103/PhysRevD.98.023013} {\bibfield  {journal} {\bibinfo  {journal} {Phys. Rev. D}\ }\textbf {\bibinfo {volume} {98}},\ \bibinfo {pages} {023013} (\bibinfo {year} {2018})},\ \Eprint {https://arxiv.org/abs/1803.09734} {arXiv:1803.09734 [astro-ph.CO]} \BibitemShut {NoStop}%
\bibitem [{\citenamefont {Rahimieh}\ \emph {et~al.}(2025{\natexlab{a}})\citenamefont {Rahimieh}, \citenamefont {Parashari}, \citenamefont {An}, \citenamefont {Driskell}, \citenamefont {Mirocha},\ and\ \citenamefont {Gluscevic}}]{Rahimieh_2025_sensitivity}%
  \BibitemOpen
  \bibfield  {author} {\bibinfo {author} {\bibfnamefont {A.}~\bibnamefont {Rahimieh}}, \bibinfo {author} {\bibfnamefont {P.}~\bibnamefont {Parashari}}, \bibinfo {author} {\bibfnamefont {R.}~\bibnamefont {An}}, \bibinfo {author} {\bibfnamefont {T.}~\bibnamefont {Driskell}}, \bibinfo {author} {\bibfnamefont {J.}~\bibnamefont {Mirocha}},\ and\ \bibinfo {author} {\bibfnamefont {V.}~\bibnamefont {Gluscevic}},\ }\bibfield  {title} {\bibinfo {title} {{Sensitivity of the Global 21-cm Signal to Dark Matter-Baryon Scattering}},\ }\href@noop {} {\  (\bibinfo {year} {2025}{\natexlab{a}})},\ \Eprint {https://arxiv.org/abs/2505.03148} {arXiv:2505.03148 [astro-ph.CO]} \BibitemShut {NoStop}%
\bibitem [{\citenamefont {Rahimieh}\ \emph {et~al.}(2025{\natexlab{b}})\citenamefont {Rahimieh}, \citenamefont {Parashari},\ and\ \citenamefont {Gluscevic}}]{Rahimieh_2025_Forecasting}%
  \BibitemOpen
  \bibfield  {author} {\bibinfo {author} {\bibfnamefont {A.}~\bibnamefont {Rahimieh}}, \bibinfo {author} {\bibfnamefont {P.}~\bibnamefont {Parashari}},\ and\ \bibinfo {author} {\bibfnamefont {V.}~\bibnamefont {Gluscevic}},\ }\bibfield  {title} {\bibinfo {title} {{Forecasting 21-cm power spectrum sensitivity to dark Matter-baryon scattering}}\ }\href {https://doi.org/10.1093/mnras/staf1326} {10.1093/mnras/staf1326} (\bibinfo {year} {2025}{\natexlab{b}}),\ \Eprint {https://arxiv.org/abs/2508.20507} {arXiv:2508.20507 [astro-ph.CO]} \BibitemShut {NoStop}%
\bibitem [{\citenamefont {Kovetz}\ \emph {et~al.}(2018)\citenamefont {Kovetz}, \citenamefont {Poulin}, \citenamefont {Gluscevic}, \citenamefont {Boddy}, \citenamefont {Barkana},\ and\ \citenamefont {Kamionkowski}}]{Kovetz_2018_Tighter}%
  \BibitemOpen
  \bibfield  {author} {\bibinfo {author} {\bibfnamefont {E.~D.}\ \bibnamefont {Kovetz}}, \bibinfo {author} {\bibfnamefont {V.}~\bibnamefont {Poulin}}, \bibinfo {author} {\bibfnamefont {V.}~\bibnamefont {Gluscevic}}, \bibinfo {author} {\bibfnamefont {K.~K.}\ \bibnamefont {Boddy}}, \bibinfo {author} {\bibfnamefont {R.}~\bibnamefont {Barkana}},\ and\ \bibinfo {author} {\bibfnamefont {M.}~\bibnamefont {Kamionkowski}},\ }\bibfield  {title} {\bibinfo {title} {{Tighter limits on dark matter explanations of the anomalous EDGES 21 cm signal}},\ }\href {https://doi.org/10.1103/PhysRevD.98.103529} {\bibfield  {journal} {\bibinfo  {journal} {Phys. Rev. D}\ }\textbf {\bibinfo {volume} {98}},\ \bibinfo {pages} {103529} (\bibinfo {year} {2018})},\ \Eprint {https://arxiv.org/abs/1807.11482} {arXiv:1807.11482 [astro-ph.CO]} \BibitemShut {NoStop}%
\bibitem [{\citenamefont {Short}\ \emph {et~al.}(2022)\citenamefont {Short}, \citenamefont {Bernal}, \citenamefont {Boddy}, \citenamefont {Gluscevic},\ and\ \citenamefont {Verde}}]{Short_2022_Dark}%
  \BibitemOpen
  \bibfield  {author} {\bibinfo {author} {\bibfnamefont {K.}~\bibnamefont {Short}}, \bibinfo {author} {\bibfnamefont {J.~L.}\ \bibnamefont {Bernal}}, \bibinfo {author} {\bibfnamefont {K.~K.}\ \bibnamefont {Boddy}}, \bibinfo {author} {\bibfnamefont {V.}~\bibnamefont {Gluscevic}},\ and\ \bibinfo {author} {\bibfnamefont {L.}~\bibnamefont {Verde}},\ }\bibfield  {title} {\bibinfo {title} {{Dark matter-baryon scattering effects on temperature perturbations and implications for cosmic dawn}},\ }\href@noop {} {\  (\bibinfo {year} {2022})},\ \Eprint {https://arxiv.org/abs/2203.16524} {arXiv:2203.16524 [astro-ph.CO]} \BibitemShut {NoStop}%
\bibitem [{\citenamefont {Driskell}\ \emph {et~al.}(2022)\citenamefont {Driskell}, \citenamefont {Nadler}, \citenamefont {Mirocha}, \citenamefont {Benson}, \citenamefont {Boddy}, \citenamefont {Morton}, \citenamefont {Lashner}, \citenamefont {An},\ and\ \citenamefont {Gluscevic}}]{Driskell_2022_Structure}%
  \BibitemOpen
  \bibfield  {author} {\bibinfo {author} {\bibfnamefont {T.}~\bibnamefont {Driskell}}, \bibinfo {author} {\bibfnamefont {E.~O.}\ \bibnamefont {Nadler}}, \bibinfo {author} {\bibfnamefont {J.}~\bibnamefont {Mirocha}}, \bibinfo {author} {\bibfnamefont {A.}~\bibnamefont {Benson}}, \bibinfo {author} {\bibfnamefont {K.~K.}\ \bibnamefont {Boddy}}, \bibinfo {author} {\bibfnamefont {T.~D.}\ \bibnamefont {Morton}}, \bibinfo {author} {\bibfnamefont {J.}~\bibnamefont {Lashner}}, \bibinfo {author} {\bibfnamefont {R.}~\bibnamefont {An}},\ and\ \bibinfo {author} {\bibfnamefont {V.}~\bibnamefont {Gluscevic}},\ }\bibfield  {title} {\bibinfo {title} {{Structure formation and the global 21-cm signal in the presence of Coulomb-like dark matter-baryon interactions}},\ }\href {https://doi.org/10.1103/PhysRevD.106.103525} {\bibfield  {journal} {\bibinfo  {journal} {Phys. Rev. D}\ }\textbf {\bibinfo {volume} {106}},\ \bibinfo {pages} {103525} (\bibinfo {year} {2022})},\ \Eprint {https://arxiv.org/abs/2209.04499} {arXiv:2209.04499
  [astro-ph.CO]} \BibitemShut {NoStop}%
\bibitem [{\citenamefont {Zhang}\ \emph {et~al.}(2024)\citenamefont {Zhang}, \citenamefont {Zu}, \citenamefont {Chen}, \citenamefont {Tsai},\ and\ \citenamefont {Fan}}]{Zhang_2024_weak}%
  \BibitemOpen
  \bibfield  {author} {\bibinfo {author} {\bibfnamefont {C.}~\bibnamefont {Zhang}}, \bibinfo {author} {\bibfnamefont {L.}~\bibnamefont {Zu}}, \bibinfo {author} {\bibfnamefont {H.-Z.}\ \bibnamefont {Chen}}, \bibinfo {author} {\bibfnamefont {Y.-L.~S.}\ \bibnamefont {Tsai}},\ and\ \bibinfo {author} {\bibfnamefont {Y.-Z.}\ \bibnamefont {Fan}},\ }\bibfield  {title} {\bibinfo {title} {{Weak lensing constraints on dark matter-baryon interactions with {\ensuremath{\mathsf{N}}}-body simulations and machine learning}},\ }\href {https://doi.org/10.1088/1475-7516/2024/08/003} {\bibfield  {journal} {\bibinfo  {journal} {JCAP}\ }\textbf {\bibinfo {volume} {08}},\ \bibinfo {pages} {003}},\ \Eprint {https://arxiv.org/abs/2402.18880} {arXiv:2402.18880 [astro-ph.CO]} \BibitemShut {NoStop}%
\bibitem [{\citenamefont {He}\ \emph {et~al.}(2025)\citenamefont {He}, \citenamefont {Ivanov}, \citenamefont {An}, \citenamefont {Driskell},\ and\ \citenamefont {Gluscevic}}]{He_2025_bounds}%
  \BibitemOpen
  \bibfield  {author} {\bibinfo {author} {\bibfnamefont {A.}~\bibnamefont {He}}, \bibinfo {author} {\bibfnamefont {M.~M.}\ \bibnamefont {Ivanov}}, \bibinfo {author} {\bibfnamefont {R.}~\bibnamefont {An}}, \bibinfo {author} {\bibfnamefont {T.}~\bibnamefont {Driskell}},\ and\ \bibinfo {author} {\bibfnamefont {V.}~\bibnamefont {Gluscevic}},\ }\bibfield  {title} {\bibinfo {title} {{Bounds on velocity-dependent dark matter-baryon scattering from large-scale structure}},\ }\href {https://doi.org/10.1088/1475-7516/2025/05/087} {\bibfield  {journal} {\bibinfo  {journal} {JCAP}\ }\textbf {\bibinfo {volume} {05}},\ \bibinfo {pages} {087}},\ \Eprint {https://arxiv.org/abs/2502.02636} {arXiv:2502.02636 [astro-ph.CO]} \BibitemShut {NoStop}%
\bibitem [{\citenamefont {Nadler}\ \emph {et~al.}(2019)\citenamefont {Nadler}, \citenamefont {Gluscevic}, \citenamefont {Boddy},\ and\ \citenamefont {Wechsler}}]{Nadler_2019_Constraints}%
  \BibitemOpen
  \bibfield  {author} {\bibinfo {author} {\bibfnamefont {E.~O.}\ \bibnamefont {Nadler}}, \bibinfo {author} {\bibfnamefont {V.}~\bibnamefont {Gluscevic}}, \bibinfo {author} {\bibfnamefont {K.~K.}\ \bibnamefont {Boddy}},\ and\ \bibinfo {author} {\bibfnamefont {R.~H.}\ \bibnamefont {Wechsler}},\ }\bibfield  {title} {\bibinfo {title} {{Constraints on Dark Matter Microphysics from the Milky Way Satellite Population}},\ }\href {https://doi.org/10.3847/2041-8213/ab1eb2} {\bibfield  {journal} {\bibinfo  {journal} {Astrophys. J. Lett.}\ }\textbf {\bibinfo {volume} {878}},\ \bibinfo {pages} {32} (\bibinfo {year} {2019})},\ \bibinfo {note} {[Erratum: Astrophys.J.Lett. 897, L46 (2020), Erratum: Astrophys.J. 897, L46 (2020)]},\ \Eprint {https://arxiv.org/abs/1904.10000} {arXiv:1904.10000 [astro-ph.CO]} \BibitemShut {NoStop}%
\bibitem [{\citenamefont {Maamari}\ \emph {et~al.}(2021)\citenamefont {Maamari}, \citenamefont {Gluscevic}, \citenamefont {Boddy}, \citenamefont {Nadler},\ and\ \citenamefont {Wechsler}}]{Maamari_2020_Bounds}%
  \BibitemOpen
  \bibfield  {author} {\bibinfo {author} {\bibfnamefont {K.}~\bibnamefont {Maamari}}, \bibinfo {author} {\bibfnamefont {V.}~\bibnamefont {Gluscevic}}, \bibinfo {author} {\bibfnamefont {K.~K.}\ \bibnamefont {Boddy}}, \bibinfo {author} {\bibfnamefont {E.~O.}\ \bibnamefont {Nadler}},\ and\ \bibinfo {author} {\bibfnamefont {R.~H.}\ \bibnamefont {Wechsler}},\ }\bibfield  {title} {\bibinfo {title} {{Bounds on velocity-dependent dark matter-proton scattering from Milky Way satellite abundance}},\ }\href {https://doi.org/10.3847/2041-8213/abd807} {\bibfield  {journal} {\bibinfo  {journal} {Astrophys. J. Lett.}\ }\textbf {\bibinfo {volume} {907}},\ \bibinfo {pages} {L46} (\bibinfo {year} {2021})},\ \Eprint {https://arxiv.org/abs/2010.02936} {arXiv:2010.02936 [astro-ph.CO]} \BibitemShut {NoStop}%
\bibitem [{\citenamefont {Nadler}\ \emph {et~al.}(2025)\citenamefont {Nadler}, \citenamefont {An}, \citenamefont {Gluscevic}, \citenamefont {Benson},\ and\ \citenamefont {Du}}]{Nadler_2025_COZMIC}%
  \BibitemOpen
  \bibfield  {author} {\bibinfo {author} {\bibfnamefont {E.~O.}\ \bibnamefont {Nadler}}, \bibinfo {author} {\bibfnamefont {R.}~\bibnamefont {An}}, \bibinfo {author} {\bibfnamefont {V.}~\bibnamefont {Gluscevic}}, \bibinfo {author} {\bibfnamefont {A.}~\bibnamefont {Benson}},\ and\ \bibinfo {author} {\bibfnamefont {X.}~\bibnamefont {Du}},\ }\bibfield  {title} {\bibinfo {title} {{COZMIC. I. Cosmological Zoom-in Simulations with Initial Conditions Beyond Cold Dark Matter}},\ }\href {https://doi.org/10.3847/1538-4357/adceef} {\bibfield  {journal} {\bibinfo  {journal} {Astrophys. J.}\ }\textbf {\bibinfo {volume} {986}},\ \bibinfo {pages} {127} (\bibinfo {year} {2025})},\ \Eprint {https://arxiv.org/abs/2410.03635} {arXiv:2410.03635 [astro-ph.CO]} \BibitemShut {NoStop}%
\bibitem [{\citenamefont {Lazare}\ \emph {et~al.}(2025)\citenamefont {Lazare}, \citenamefont {Kovetz}, \citenamefont {Boddy},\ and\ \citenamefont {Munoz}}]{Lazare_2025_First}%
  \BibitemOpen
  \bibfield  {author} {\bibinfo {author} {\bibfnamefont {H.}~\bibnamefont {Lazare}}, \bibinfo {author} {\bibfnamefont {E.~D.}\ \bibnamefont {Kovetz}}, \bibinfo {author} {\bibfnamefont {K.~K.}\ \bibnamefont {Boddy}},\ and\ \bibinfo {author} {\bibfnamefont {J.~B.}\ \bibnamefont {Munoz}},\ }\bibfield  {title} {\bibinfo {title} {{First galaxy ultraviolet luminosity function limits on dark matter-proton scattering}},\ }\href@noop {} {\  (\bibinfo {year} {2025})},\ \Eprint {https://arxiv.org/abs/2510.10757} {arXiv:2510.10757 [astro-ph.CO]} \BibitemShut {NoStop}%
\bibitem [{\citenamefont {Das}\ \emph {et~al.}(2026)\citenamefont {Das}, \citenamefont {Mondol}, \citenamefont {Singh},\ and\ \citenamefont {Laha}}]{Das:2025bnr}%
  \BibitemOpen
  \bibfield  {author} {\bibinfo {author} {\bibfnamefont {S.}~\bibnamefont {Das}}, \bibinfo {author} {\bibfnamefont {R.}~\bibnamefont {Mondol}}, \bibinfo {author} {\bibfnamefont {A.}~\bibnamefont {Singh}},\ and\ \bibinfo {author} {\bibfnamefont {R.}~\bibnamefont {Laha}},\ }\bibfield  {title} {\bibinfo {title} {{Dark Secrets of Baryons: Illuminating Dark Matter-Baryon Interactions with JWST}},\ }\href {https://doi.org/10.1103/44nx-sgk4} {\bibfield  {journal} {\bibinfo  {journal} {Phys. Rev. Lett.}\ }\textbf {\bibinfo {volume} {137}},\ \bibinfo {pages} {011002} (\bibinfo {year} {2026})},\ \Eprint {https://arxiv.org/abs/2511.02906} {arXiv:2511.02906 [hep-ph]} \BibitemShut {NoStop}%
\bibitem [{\citenamefont {He}\ \emph {et~al.}(2023)\citenamefont {He}, \citenamefont {Ivanov}, \citenamefont {An},\ and\ \citenamefont {Gluscevic}}]{He_2023_S8}%
  \BibitemOpen
  \bibfield  {author} {\bibinfo {author} {\bibfnamefont {A.}~\bibnamefont {He}}, \bibinfo {author} {\bibfnamefont {M.~M.}\ \bibnamefont {Ivanov}}, \bibinfo {author} {\bibfnamefont {R.}~\bibnamefont {An}},\ and\ \bibinfo {author} {\bibfnamefont {V.}~\bibnamefont {Gluscevic}},\ }\bibfield  {title} {\bibinfo {title} {{S$_{8}$ Tension in the Context of Dark Matter{\textendash}Baryon Scattering}},\ }\href {https://doi.org/10.3847/2041-8213/acdb63} {\bibfield  {journal} {\bibinfo  {journal} {Astrophys. J. Lett.}\ }\textbf {\bibinfo {volume} {954}},\ \bibinfo {pages} {L8} (\bibinfo {year} {2023})},\ \Eprint {https://arxiv.org/abs/2301.08260} {arXiv:2301.08260 [astro-ph.CO]} \BibitemShut {NoStop}%
\bibitem [{\citenamefont {Smith}\ \emph {et~al.}(2021)\citenamefont {Smith}, \citenamefont {Poulin}, \citenamefont {Bernal}, \citenamefont {Boddy}, \citenamefont {Kamionkowski},\ and\ \citenamefont {Murgia}}]{Smith_2020_Early}%
  \BibitemOpen
  \bibfield  {author} {\bibinfo {author} {\bibfnamefont {T.~L.}\ \bibnamefont {Smith}}, \bibinfo {author} {\bibfnamefont {V.}~\bibnamefont {Poulin}}, \bibinfo {author} {\bibfnamefont {J.~L.}\ \bibnamefont {Bernal}}, \bibinfo {author} {\bibfnamefont {K.~K.}\ \bibnamefont {Boddy}}, \bibinfo {author} {\bibfnamefont {M.}~\bibnamefont {Kamionkowski}},\ and\ \bibinfo {author} {\bibfnamefont {R.}~\bibnamefont {Murgia}},\ }\bibfield  {title} {\bibinfo {title} {{Early dark energy is not excluded by current large-scale structure data}},\ }\href {https://doi.org/10.1103/PhysRevD.103.123542} {\bibfield  {journal} {\bibinfo  {journal} {Phys. Rev. D}\ }\textbf {\bibinfo {volume} {103}},\ \bibinfo {pages} {123542} (\bibinfo {year} {2021})},\ \Eprint {https://arxiv.org/abs/2009.10740} {arXiv:2009.10740 [astro-ph.CO]} \BibitemShut {NoStop}%
\bibitem [{\citenamefont {Herold}\ \emph {et~al.}(2022)\citenamefont {Herold}, \citenamefont {Ferreira},\ and\ \citenamefont {Komatsu}}]{Herold_2021_New}%
  \BibitemOpen
  \bibfield  {author} {\bibinfo {author} {\bibfnamefont {L.}~\bibnamefont {Herold}}, \bibinfo {author} {\bibfnamefont {E.~G.~M.}\ \bibnamefont {Ferreira}},\ and\ \bibinfo {author} {\bibfnamefont {E.}~\bibnamefont {Komatsu}},\ }\bibfield  {title} {\bibinfo {title} {{New Constraint on Early Dark Energy from Planck and BOSS Data Using the Profile Likelihood}},\ }\href {https://doi.org/10.3847/2041-8213/ac63a3} {\bibfield  {journal} {\bibinfo  {journal} {Astrophys. J. Lett.}\ }\textbf {\bibinfo {volume} {929}},\ \bibinfo {pages} {L16} (\bibinfo {year} {2022})},\ \Eprint {https://arxiv.org/abs/2112.12140} {arXiv:2112.12140 [astro-ph.CO]} \BibitemShut {NoStop}%
\bibitem [{\citenamefont {Herold}\ and\ \citenamefont {Ferreira}(2023)}]{Herold:2022iib}%
  \BibitemOpen
  \bibfield  {author} {\bibinfo {author} {\bibfnamefont {L.}~\bibnamefont {Herold}}\ and\ \bibinfo {author} {\bibfnamefont {E.~G.~M.}\ \bibnamefont {Ferreira}},\ }\bibfield  {title} {\bibinfo {title} {{Resolving the Hubble tension with early dark energy}},\ }\href {https://doi.org/10.1103/PhysRevD.108.043513} {\bibfield  {journal} {\bibinfo  {journal} {Phys. Rev. D}\ }\textbf {\bibinfo {volume} {108}},\ \bibinfo {pages} {043513} (\bibinfo {year} {2023})},\ \Eprint {https://arxiv.org/abs/2210.16296} {arXiv:2210.16296 [astro-ph.CO]} \BibitemShut {NoStop}%
\bibitem [{\citenamefont {Karwal}\ \emph {et~al.}(2024)\citenamefont {Karwal}, \citenamefont {Patel}, \citenamefont {Bartlett}, \citenamefont {Poulin}, \citenamefont {Smith},\ and\ \citenamefont {Pfeffer}}]{Karwal_2024_Procoli}%
  \BibitemOpen
  \bibfield  {author} {\bibinfo {author} {\bibfnamefont {T.}~\bibnamefont {Karwal}}, \bibinfo {author} {\bibfnamefont {Y.}~\bibnamefont {Patel}}, \bibinfo {author} {\bibfnamefont {A.}~\bibnamefont {Bartlett}}, \bibinfo {author} {\bibfnamefont {V.}~\bibnamefont {Poulin}}, \bibinfo {author} {\bibfnamefont {T.~L.}\ \bibnamefont {Smith}},\ and\ \bibinfo {author} {\bibfnamefont {D.~N.}\ \bibnamefont {Pfeffer}},\ }\bibfield  {title} {\bibinfo {title} {{Procoli: Profiles of cosmological likelihoods}},\ }\href@noop {} {\  (\bibinfo {year} {2024})},\ \Eprint {https://arxiv.org/abs/2401.14225} {arXiv:2401.14225 [astro-ph.CO]} \BibitemShut {NoStop}%
\bibitem [{\citenamefont {Holm}\ \emph {et~al.}(2023)\citenamefont {Holm}, \citenamefont {Herold}, \citenamefont {Simon}, \citenamefont {Ferreira}, \citenamefont {Hannestad}, \citenamefont {Poulin},\ and\ \citenamefont {Tram}}]{Holm:2023laa}%
  \BibitemOpen
  \bibfield  {author} {\bibinfo {author} {\bibfnamefont {E.~B.}\ \bibnamefont {Holm}}, \bibinfo {author} {\bibfnamefont {L.}~\bibnamefont {Herold}}, \bibinfo {author} {\bibfnamefont {T.}~\bibnamefont {Simon}}, \bibinfo {author} {\bibfnamefont {E.~G.~M.}\ \bibnamefont {Ferreira}}, \bibinfo {author} {\bibfnamefont {S.}~\bibnamefont {Hannestad}}, \bibinfo {author} {\bibfnamefont {V.}~\bibnamefont {Poulin}},\ and\ \bibinfo {author} {\bibfnamefont {T.}~\bibnamefont {Tram}},\ }\bibfield  {title} {\bibinfo {title} {{Bayesian and frequentist investigation of prior effects in EFT of LSS analyses of full-shape BOSS and eBOSS data}},\ }\href {https://doi.org/10.1103/PhysRevD.108.123514} {\bibfield  {journal} {\bibinfo  {journal} {Phys. Rev. D}\ }\textbf {\bibinfo {volume} {108}},\ \bibinfo {pages} {123514} (\bibinfo {year} {2023})},\ \Eprint {https://arxiv.org/abs/2309.04468} {arXiv:2309.04468 [astro-ph.CO]} \BibitemShut {NoStop}%
\bibitem [{\citenamefont {Zhang}\ \emph {et~al.}(2025)\citenamefont {Zhang}, \citenamefont {Bonici}, \citenamefont {D'Amico}, \citenamefont {Paradiso},\ and\ \citenamefont {Percival}}]{Zhang:2024thl}%
  \BibitemOpen
  \bibfield  {author} {\bibinfo {author} {\bibfnamefont {H.}~\bibnamefont {Zhang}}, \bibinfo {author} {\bibfnamefont {M.}~\bibnamefont {Bonici}}, \bibinfo {author} {\bibfnamefont {G.}~\bibnamefont {D'Amico}}, \bibinfo {author} {\bibfnamefont {S.}~\bibnamefont {Paradiso}},\ and\ \bibinfo {author} {\bibfnamefont {W.~J.}\ \bibnamefont {Percival}},\ }\bibfield  {title} {\bibinfo {title} {{HOD-informed prior for EFT-based full-shape analyses of LSS}},\ }\href {https://doi.org/10.1088/1475-7516/2025/04/041} {\bibfield  {journal} {\bibinfo  {journal} {JCAP}\ }\textbf {\bibinfo {volume} {04}},\ \bibinfo {pages} {041}},\ \Eprint {https://arxiv.org/abs/2409.12937} {arXiv:2409.12937 [astro-ph.CO]} \BibitemShut {NoStop}%
\bibitem [{\citenamefont {Chudaykin}\ \emph {et~al.}(2026)\citenamefont {Chudaykin}, \citenamefont {Ivanov},\ and\ \citenamefont {Nishimichi}}]{Chudaykin:2024wlw}%
  \BibitemOpen
  \bibfield  {author} {\bibinfo {author} {\bibfnamefont {A.}~\bibnamefont {Chudaykin}}, \bibinfo {author} {\bibfnamefont {M.~M.}\ \bibnamefont {Ivanov}},\ and\ \bibinfo {author} {\bibfnamefont {T.}~\bibnamefont {Nishimichi}},\ }\bibfield  {title} {\bibinfo {title} {{Priors and scale cuts in EFT-based full-shape analyses}},\ }\href {https://doi.org/10.1103/glg2-py5y} {\bibfield  {journal} {\bibinfo  {journal} {Phys. Rev. D}\ }\textbf {\bibinfo {volume} {113}},\ \bibinfo {pages} {063524} (\bibinfo {year} {2026})},\ \Eprint {https://arxiv.org/abs/2410.16358} {arXiv:2410.16358 [astro-ph.CO]} \BibitemShut {NoStop}%
\bibitem [{\citenamefont {Secco}\ \emph {et~al.}(2022)\citenamefont {Secco} \emph {et~al.}}]{DES:2021vln}%
  \BibitemOpen
  \bibfield  {author} {\bibinfo {author} {\bibfnamefont {L.~F.}\ \bibnamefont {Secco}} \emph {et~al.} (\bibinfo {collaboration} {DES}),\ }\bibfield  {title} {\bibinfo {title} {{Dark Energy Survey Year 3 results: Cosmology from cosmic shear and robustness to modeling uncertainty}},\ }\href {https://doi.org/10.1103/PhysRevD.105.023515} {\bibfield  {journal} {\bibinfo  {journal} {Phys. Rev. D}\ }\textbf {\bibinfo {volume} {105}},\ \bibinfo {pages} {023515} (\bibinfo {year} {2022})},\ \Eprint {https://arxiv.org/abs/2105.13544} {arXiv:2105.13544 [astro-ph.CO]} \BibitemShut {NoStop}%
\bibitem [{\citenamefont {Abbott}\ \emph {et~al.}(2023)\citenamefont {Abbott} \emph {et~al.}}]{Kilo-DegreeSurvey:2023gfr}%
  \BibitemOpen
  \bibfield  {author} {\bibinfo {author} {\bibfnamefont {T.~M.~C.}\ \bibnamefont {Abbott}} \emph {et~al.} (\bibinfo {collaboration} {Kilo-Degree Survey, DES}),\ }\bibfield  {title} {\bibinfo {title} {{DES Y3 + KiDS-1000: Consistent cosmology combining cosmic shear surveys}},\ }\href {https://doi.org/10.21105/astro.2305.17173} {\bibfield  {journal} {\bibinfo  {journal} {Open J. Astrophys.}\ }\textbf {\bibinfo {volume} {6}},\ \bibinfo {pages} {2305.17173} (\bibinfo {year} {2023})},\ \Eprint {https://arxiv.org/abs/2305.17173} {arXiv:2305.17173 [astro-ph.CO]} \BibitemShut {NoStop}%
\bibitem [{\citenamefont {Aric{\`o}}\ \emph {et~al.}(2023)\citenamefont {Aric{\`o}}, \citenamefont {Angulo}, \citenamefont {Zennaro}, \citenamefont {Contreras}, \citenamefont {Chen},\ and\ \citenamefont {Hern{\'a}ndez-Monteagudo}}]{Arico:2023ocu}%
  \BibitemOpen
  \bibfield  {author} {\bibinfo {author} {\bibfnamefont {G.}~\bibnamefont {Aric{\`o}}}, \bibinfo {author} {\bibfnamefont {R.~E.}\ \bibnamefont {Angulo}}, \bibinfo {author} {\bibfnamefont {M.}~\bibnamefont {Zennaro}}, \bibinfo {author} {\bibfnamefont {S.}~\bibnamefont {Contreras}}, \bibinfo {author} {\bibfnamefont {A.}~\bibnamefont {Chen}},\ and\ \bibinfo {author} {\bibfnamefont {C.}~\bibnamefont {Hern{\'a}ndez-Monteagudo}},\ }\bibfield  {title} {\bibinfo {title} {{DES Y3 cosmic shear down to small scales: Constraints on cosmology and baryons}},\ }\href {https://doi.org/10.1051/0004-6361/202346539} {\bibfield  {journal} {\bibinfo  {journal} {Astron. Astrophys.}\ }\textbf {\bibinfo {volume} {678}},\ \bibinfo {pages} {A109} (\bibinfo {year} {2023})},\ \Eprint {https://arxiv.org/abs/2303.05537} {arXiv:2303.05537 [astro-ph.CO]} \BibitemShut {NoStop}%
\bibitem [{\citenamefont {Fowlie}(2024)}]{Fowlie_2024_Bayes}%
  \BibitemOpen
  \bibfield  {author} {\bibinfo {author} {\bibfnamefont {A.}~\bibnamefont {Fowlie}},\ }\bibfield  {title} {\bibinfo {title} {{The Bayes factor surface for searches for new physics}},\ }\href {https://doi.org/10.1140/epjc/s10052-024-12792-9} {\bibfield  {journal} {\bibinfo  {journal} {Eur. Phys. J. C}\ }\textbf {\bibinfo {volume} {84}},\ \bibinfo {pages} {426} (\bibinfo {year} {2024})},\ \Eprint {https://arxiv.org/abs/2401.11710} {arXiv:2401.11710 [hep-ph]} \BibitemShut {NoStop}%
\bibitem [{\citenamefont {Herold}\ \emph {et~al.}(2025)\citenamefont {Herold}, \citenamefont {Ferreira},\ and\ \citenamefont {Heinrich}}]{Herold_2024_profile}%
  \BibitemOpen
  \bibfield  {author} {\bibinfo {author} {\bibfnamefont {L.}~\bibnamefont {Herold}}, \bibinfo {author} {\bibfnamefont {E.~G.~M.}\ \bibnamefont {Ferreira}},\ and\ \bibinfo {author} {\bibfnamefont {L.}~\bibnamefont {Heinrich}},\ }\bibfield  {title} {\bibinfo {title} {{Profile likelihoods in cosmology: When, why, and how illustrated with {\ensuremath{\Lambda}}CDM, massive neutrinos, and dark energy}},\ }\href {https://doi.org/10.1103/PhysRevD.111.083504} {\bibfield  {journal} {\bibinfo  {journal} {Phys. Rev. D}\ }\textbf {\bibinfo {volume} {111}},\ \bibinfo {pages} {083504} (\bibinfo {year} {2025})},\ \Eprint {https://arxiv.org/abs/2408.07700} {arXiv:2408.07700 [astro-ph.CO]} \BibitemShut {NoStop}%
\bibitem [{\citenamefont {Ma}\ and\ \citenamefont {Bertschinger}(1995)}]{Ma_1995_cosmological}%
  \BibitemOpen
  \bibfield  {author} {\bibinfo {author} {\bibfnamefont {C.-P.}\ \bibnamefont {Ma}}\ and\ \bibinfo {author} {\bibfnamefont {E.}~\bibnamefont {Bertschinger}},\ }\bibfield  {title} {\bibinfo {title} {{Cosmological perturbation theory in the synchronous and conformal Newtonian gauges}},\ }\href {https://doi.org/10.1086/176550} {\bibfield  {journal} {\bibinfo  {journal} {Astrophys. J.}\ }\textbf {\bibinfo {volume} {455}},\ \bibinfo {pages} {7} (\bibinfo {year} {1995})},\ \Eprint {https://arxiv.org/abs/astro-ph/9506072} {arXiv:astro-ph/9506072} \BibitemShut {NoStop}%
\bibitem [{\citenamefont {Ali-Ha{\"\i}moud}\ \emph {et~al.}(2024)\citenamefont {Ali-Ha{\"\i}moud}, \citenamefont {Gandhi},\ and\ \citenamefont {Smith}}]{Ali-Haimoud_2024_exact}%
  \BibitemOpen
  \bibfield  {author} {\bibinfo {author} {\bibfnamefont {Y.}~\bibnamefont {Ali-Ha{\"\i}moud}}, \bibinfo {author} {\bibfnamefont {S.~S.}\ \bibnamefont {Gandhi}},\ and\ \bibinfo {author} {\bibfnamefont {T.~L.}\ \bibnamefont {Smith}},\ }\bibfield  {title} {\bibinfo {title} {{Exact treatment of weak dark matter-baryon scattering for linear-cosmology observables}},\ }\href {https://doi.org/10.1103/PhysRevD.109.083523} {\bibfield  {journal} {\bibinfo  {journal} {Phys. Rev. D}\ }\textbf {\bibinfo {volume} {109}},\ \bibinfo {pages} {083523} (\bibinfo {year} {2024})},\ \Eprint {https://arxiv.org/abs/2312.08497} {arXiv:2312.08497 [astro-ph.CO]} \BibitemShut {NoStop}%
\bibitem [{\citenamefont {Neyman}(1937)}]{Neyman_1937_Outline}%
  \BibitemOpen
  \bibfield  {author} {\bibinfo {author} {\bibfnamefont {J.}~\bibnamefont {Neyman}},\ }\bibfield  {title} {\bibinfo {title} {{Outline of a Theory of Statistical Estimation Based on the Classical Theory of Probability}},\ }\href {https://doi.org/10.1098/rsta.1937.0005} {\bibfield  {journal} {\bibinfo  {journal} {Phil. Trans. Roy. Soc. Lond. A}\ }\textbf {\bibinfo {volume} {236}},\ \bibinfo {pages} {333} (\bibinfo {year} {1937})}\BibitemShut {NoStop}%
\bibitem [{\citenamefont {Wilks}(1938)}]{Wilks_1938_Large-Sample}%
  \BibitemOpen
  \bibfield  {author} {\bibinfo {author} {\bibfnamefont {S.~S.}\ \bibnamefont {Wilks}},\ }\bibfield  {title} {\bibinfo {title} {{The Large-Sample Distribution of the Likelihood Ratio for Testing Composite Hypotheses}},\ }\href {https://doi.org/10.1214/aoms/1177732360} {\bibfield  {journal} {\bibinfo  {journal} {Annals Math. Statist.}\ }\textbf {\bibinfo {volume} {9}},\ \bibinfo {pages} {60} (\bibinfo {year} {1938})}\BibitemShut {NoStop}%
\bibitem [{\citenamefont {Feldman}\ and\ \citenamefont {Cousins}(1998)}]{Feldman_1997_Unified}%
  \BibitemOpen
  \bibfield  {author} {\bibinfo {author} {\bibfnamefont {G.~J.}\ \bibnamefont {Feldman}}\ and\ \bibinfo {author} {\bibfnamefont {R.~D.}\ \bibnamefont {Cousins}},\ }\bibfield  {title} {\bibinfo {title} {{A Unified approach to the classical statistical analysis of small signals}},\ }\href {https://doi.org/10.1103/PhysRevD.57.3873} {\bibfield  {journal} {\bibinfo  {journal} {Phys. Rev. D}\ }\textbf {\bibinfo {volume} {57}},\ \bibinfo {pages} {3873} (\bibinfo {year} {1998})},\ \Eprint {https://arxiv.org/abs/physics/9711021} {arXiv:physics/9711021} \BibitemShut {NoStop}%
\bibitem [{\citenamefont {Blas}\ \emph {et~al.}(2011)\citenamefont {Blas}, \citenamefont {Lesgourgues},\ and\ \citenamefont {Tram}}]{Blas_2011_CLASS}%
  \BibitemOpen
  \bibfield  {author} {\bibinfo {author} {\bibfnamefont {D.}~\bibnamefont {Blas}}, \bibinfo {author} {\bibfnamefont {J.}~\bibnamefont {Lesgourgues}},\ and\ \bibinfo {author} {\bibfnamefont {T.}~\bibnamefont {Tram}},\ }\bibfield  {title} {\bibinfo {title} {{The Cosmic Linear Anisotropy Solving System (CLASS) II: Approximation schemes}},\ }\href {https://doi.org/10.1088/1475-7516/2011/07/034} {\bibfield  {journal} {\bibinfo  {journal} {JCAP}\ }\textbf {\bibinfo {volume} {07}},\ \bibinfo {pages} {034}},\ \Eprint {https://arxiv.org/abs/1104.2933} {arXiv:1104.2933 [astro-ph.CO]} \BibitemShut {NoStop}%
\bibitem [{\citenamefont {Torrado}\ and\ \citenamefont {Lewis}(2021)}]{Torrado_2020_Cobaya}%
  \BibitemOpen
  \bibfield  {author} {\bibinfo {author} {\bibfnamefont {J.}~\bibnamefont {Torrado}}\ and\ \bibinfo {author} {\bibfnamefont {A.}~\bibnamefont {Lewis}},\ }\bibfield  {title} {\bibinfo {title} {{Cobaya: Code for Bayesian Analysis of hierarchical physical models}},\ }\href {https://doi.org/10.1088/1475-7516/2021/05/057} {\bibfield  {journal} {\bibinfo  {journal} {JCAP}\ }\textbf {\bibinfo {volume} {05}},\ \bibinfo {pages} {057}},\ \Eprint {https://arxiv.org/abs/2005.05290} {arXiv:2005.05290 [astro-ph.IM]} \BibitemShut {NoStop}%
\bibitem [{\citenamefont {Brinckmann}\ and\ \citenamefont {Lesgourgues}(2018)}]{Brinckmann_2018_MontePython}%
  \BibitemOpen
  \bibfield  {author} {\bibinfo {author} {\bibfnamefont {T.}~\bibnamefont {Brinckmann}}\ and\ \bibinfo {author} {\bibfnamefont {J.}~\bibnamefont {Lesgourgues}},\ }\bibfield  {title} {\bibinfo {title} {{MontePython 3: boosted MCMC sampler and other features}},\ }\href@noop {} {\  (\bibinfo {year} {2018})},\ \Eprint {https://arxiv.org/abs/1804.07261} {arXiv:1804.07261 [astro-ph.CO]} \BibitemShut {NoStop}%
%%CITATION = ARXIV:1804.07261;%%
\bibitem [{\citenamefont {Karwal}\ and\ \citenamefont {Kamionkowski}(2016)}]{Karwal:2016vyq}%
  \BibitemOpen
  \bibfield  {author} {\bibinfo {author} {\bibfnamefont {T.}~\bibnamefont {Karwal}}\ and\ \bibinfo {author} {\bibfnamefont {M.}~\bibnamefont {Kamionkowski}},\ }\bibfield  {title} {\bibinfo {title} {{Dark energy at early times, the Hubble parameter, and the string axiverse}},\ }\href {https://doi.org/10.1103/PhysRevD.94.103523} {\bibfield  {journal} {\bibinfo  {journal} {Phys. Rev. D}\ }\textbf {\bibinfo {volume} {94}},\ \bibinfo {pages} {103523} (\bibinfo {year} {2016})},\ \Eprint {https://arxiv.org/abs/1608.01309} {arXiv:1608.01309 [astro-ph.CO]} \BibitemShut {NoStop}%
\bibitem [{\citenamefont {Poulin}\ \emph {et~al.}(2019)\citenamefont {Poulin}, \citenamefont {Smith}, \citenamefont {Karwal},\ and\ \citenamefont {Kamionkowski}}]{Poulin_2018_Early}%
  \BibitemOpen
  \bibfield  {author} {\bibinfo {author} {\bibfnamefont {V.}~\bibnamefont {Poulin}}, \bibinfo {author} {\bibfnamefont {T.~L.}\ \bibnamefont {Smith}}, \bibinfo {author} {\bibfnamefont {T.}~\bibnamefont {Karwal}},\ and\ \bibinfo {author} {\bibfnamefont {M.}~\bibnamefont {Kamionkowski}},\ }\bibfield  {title} {\bibinfo {title} {{Early Dark Energy Can Resolve The Hubble Tension}},\ }\href {https://doi.org/10.1103/PhysRevLett.122.221301} {\bibfield  {journal} {\bibinfo  {journal} {Phys. Rev. Lett.}\ }\textbf {\bibinfo {volume} {122}},\ \bibinfo {pages} {221301} (\bibinfo {year} {2019})},\ \Eprint {https://arxiv.org/abs/1811.04083} {arXiv:1811.04083 [astro-ph.CO]} \BibitemShut {NoStop}%
\bibitem [{\citenamefont {Poulin}\ \emph {et~al.}(2023)\citenamefont {Poulin}, \citenamefont {Smith},\ and\ \citenamefont {Karwal}}]{Poulin:2023lkg}%
  \BibitemOpen
  \bibfield  {author} {\bibinfo {author} {\bibfnamefont {V.}~\bibnamefont {Poulin}}, \bibinfo {author} {\bibfnamefont {T.~L.}\ \bibnamefont {Smith}},\ and\ \bibinfo {author} {\bibfnamefont {T.}~\bibnamefont {Karwal}},\ }\bibfield  {title} {\bibinfo {title} {{The Ups and Downs of Early Dark Energy solutions to the Hubble tension: A review of models, hints and constraints circa 2023}},\ }\href {https://doi.org/10.1016/j.dark.2023.101348} {\bibfield  {journal} {\bibinfo  {journal} {Phys. Dark Univ.}\ }\textbf {\bibinfo {volume} {42}},\ \bibinfo {pages} {101348} (\bibinfo {year} {2023})},\ \Eprint {https://arxiv.org/abs/2302.09032} {arXiv:2302.09032 [astro-ph.CO]} \BibitemShut {NoStop}%
\bibitem [{\citenamefont {Kamionkowski}\ and\ \citenamefont {Riess}(2023)}]{Kamionkowski:2022pkx}%
  \BibitemOpen
  \bibfield  {author} {\bibinfo {author} {\bibfnamefont {M.}~\bibnamefont {Kamionkowski}}\ and\ \bibinfo {author} {\bibfnamefont {A.~G.}\ \bibnamefont {Riess}},\ }\bibfield  {title} {\bibinfo {title} {{The Hubble Tension and Early Dark Energy}},\ }\href {https://doi.org/10.1146/annurev-nucl-111422-024107} {\bibfield  {journal} {\bibinfo  {journal} {Ann. Rev. Nucl. Part. Sci.}\ }\textbf {\bibinfo {volume} {73}},\ \bibinfo {pages} {153} (\bibinfo {year} {2023})},\ \Eprint {https://arxiv.org/abs/2211.04492} {arXiv:2211.04492 [astro-ph.CO]} \BibitemShut {NoStop}%
\bibitem [{\citenamefont {Khalife}\ \emph {et~al.}(2026)\citenamefont {Khalife} \emph {et~al.}}]{SPT-3G:2025vyw}%
  \BibitemOpen
  \bibfield  {author} {\bibinfo {author} {\bibfnamefont {A.~R.}\ \bibnamefont {Khalife}} \emph {et~al.} (\bibinfo {collaboration} {SPT-3G}),\ }\bibfield  {title} {\bibinfo {title} {{SPT-3G D1: Axion early dark energy with CMB experiments and DESI observations}},\ }\href {https://doi.org/10.1103/8jjr-7hpb} {\bibfield  {journal} {\bibinfo  {journal} {Phys. Rev. D}\ }\textbf {\bibinfo {volume} {113}},\ \bibinfo {pages} {103546} (\bibinfo {year} {2026})},\ \Eprint {https://arxiv.org/abs/2507.23355} {arXiv:2507.23355 [astro-ph.CO]} \BibitemShut {NoStop}%
\bibitem [{\citenamefont {{Harris}}\ \emph {et~al.}(2020)\citenamefont {{Harris}}, \citenamefont {{Millman}}, \citenamefont {{van der Walt}}, \citenamefont {{Gommers}}, \citenamefont {{Virtanen}}, \citenamefont {{Cournapeau}}, \citenamefont {{Wieser}}, \citenamefont {{Taylor}}, \citenamefont {{Berg}}, \citenamefont {{Smith}}, \citenamefont {{Kern}}, \citenamefont {{Picus}}, \citenamefont {{Hoyer}}, \citenamefont {{van Kerkwijk}}, \citenamefont {{Brett}}, \citenamefont {{Haldane}}, \citenamefont {{del R{\'\i}o}}, \citenamefont {{Wiebe}}, \citenamefont {{Peterson}}, \citenamefont {{G{\'e}rard-Marchant}}, \citenamefont {{Sheppard}}, \citenamefont {{Reddy}}, \citenamefont {{Weckesser}}, \citenamefont {{Abbasi}}, \citenamefont {{Gohlke}},\ and\ \citenamefont {{Oliphant}}}]{Harris_2020_NumPy}%
  \BibitemOpen
  \bibfield  {author} {\bibinfo {author} {\bibfnamefont {C.~R.}\ \bibnamefont {{Harris}}}, \bibinfo {author} {\bibfnamefont {K.~J.}\ \bibnamefont {{Millman}}}, \bibinfo {author} {\bibfnamefont {S.~J.}\ \bibnamefont {{van der Walt}}}, \bibinfo {author} {\bibfnamefont {R.}~\bibnamefont {{Gommers}}}, \bibinfo {author} {\bibfnamefont {P.}~\bibnamefont {{Virtanen}}}, \bibinfo {author} {\bibfnamefont {D.}~\bibnamefont {{Cournapeau}}}, \bibinfo {author} {\bibfnamefont {E.}~\bibnamefont {{Wieser}}}, \bibinfo {author} {\bibfnamefont {J.}~\bibnamefont {{Taylor}}}, \bibinfo {author} {\bibfnamefont {S.}~\bibnamefont {{Berg}}}, \bibinfo {author} {\bibfnamefont {N.~J.}\ \bibnamefont {{Smith}}}, \bibinfo {author} {\bibfnamefont {R.}~\bibnamefont {{Kern}}}, \bibinfo {author} {\bibfnamefont {M.}~\bibnamefont {{Picus}}}, \bibinfo {author} {\bibfnamefont {S.}~\bibnamefont {{Hoyer}}}, \bibinfo {author} {\bibfnamefont {M.~H.}\ \bibnamefont {{van Kerkwijk}}}, \bibinfo {author} {\bibfnamefont {M.}~\bibnamefont {{Brett}}}, \bibinfo
  {author} {\bibfnamefont {A.}~\bibnamefont {{Haldane}}}, \bibinfo {author} {\bibfnamefont {J.~F.}\ \bibnamefont {{del R{\'\i}o}}}, \bibinfo {author} {\bibfnamefont {M.}~\bibnamefont {{Wiebe}}}, \bibinfo {author} {\bibfnamefont {P.}~\bibnamefont {{Peterson}}}, \bibinfo {author} {\bibfnamefont {P.}~\bibnamefont {{G{\'e}rard-Marchant}}}, \bibinfo {author} {\bibfnamefont {K.}~\bibnamefont {{Sheppard}}}, \bibinfo {author} {\bibfnamefont {T.}~\bibnamefont {{Reddy}}}, \bibinfo {author} {\bibfnamefont {W.}~\bibnamefont {{Weckesser}}}, \bibinfo {author} {\bibfnamefont {H.}~\bibnamefont {{Abbasi}}}, \bibinfo {author} {\bibfnamefont {C.}~\bibnamefont {{Gohlke}}},\ and\ \bibinfo {author} {\bibfnamefont {T.~E.}\ \bibnamefont {{Oliphant}}},\ }\bibfield  {title} {\bibinfo {title} {{Array programming with NumPy}},\ }\href {https://doi.org/10.1038/s41586-020-2649-2} {\bibfield  {journal} {\bibinfo  {journal} {\nat}\ }\textbf {\bibinfo {volume} {585}},\ \bibinfo {pages} {357} (\bibinfo {year} {2020})},\ \Eprint
  {https://arxiv.org/abs/2006.10256} {arXiv:2006.10256 [cs.MS]} \BibitemShut {NoStop}%
\bibitem [{\citenamefont {{Virtanen}}\ \emph {et~al.}(2020)\citenamefont {{Virtanen}}, \citenamefont {{Gommers}}, \citenamefont {{Oliphant}}, \citenamefont {{Haberland}}, \citenamefont {{Reddy}}, \citenamefont {{Cournapeau}}, \citenamefont {{Burovski}}, \citenamefont {{Peterson}}, \citenamefont {{Weckesser}}, \citenamefont {{Bright}}, \citenamefont {{van der Walt}}, \citenamefont {{Brett}}, \citenamefont {{Wilson}}, \citenamefont {{Millman}}, \citenamefont {{Mayorov}}, \citenamefont {{Nelson}}, \citenamefont {{Jones}}, \citenamefont {{Kern}}, \citenamefont {{Larson}}, \citenamefont {{Carey}}, \citenamefont {{Polat}}, \citenamefont {{Feng}}, \citenamefont {{Moore}}, \citenamefont {{VanderPlas}}, \citenamefont {{Laxalde}}, \citenamefont {{Perktold}}, \citenamefont {{Cimrman}}, \citenamefont {{Henriksen}}, \citenamefont {{Quintero}}, \citenamefont {{Harris}}, \citenamefont {{Archibald}}, \citenamefont {{Ribeiro}}, \citenamefont {{Pedregosa}}, \citenamefont {{van Mulbregt}},\ and\ \citenamefont {{SciPy 1. 0
  Contributors}}}]{Virtanen_2020_SciPy}%
  \BibitemOpen
  \bibfield  {author} {\bibinfo {author} {\bibfnamefont {P.}~\bibnamefont {{Virtanen}}}, \bibinfo {author} {\bibfnamefont {R.}~\bibnamefont {{Gommers}}}, \bibinfo {author} {\bibfnamefont {T.~E.}\ \bibnamefont {{Oliphant}}}, \bibinfo {author} {\bibfnamefont {M.}~\bibnamefont {{Haberland}}}, \bibinfo {author} {\bibfnamefont {T.}~\bibnamefont {{Reddy}}}, \bibinfo {author} {\bibfnamefont {D.}~\bibnamefont {{Cournapeau}}}, \bibinfo {author} {\bibfnamefont {E.}~\bibnamefont {{Burovski}}}, \bibinfo {author} {\bibfnamefont {P.}~\bibnamefont {{Peterson}}}, \bibinfo {author} {\bibfnamefont {W.}~\bibnamefont {{Weckesser}}}, \bibinfo {author} {\bibfnamefont {J.}~\bibnamefont {{Bright}}}, \bibinfo {author} {\bibfnamefont {S.~J.}\ \bibnamefont {{van der Walt}}}, \bibinfo {author} {\bibfnamefont {M.}~\bibnamefont {{Brett}}}, \bibinfo {author} {\bibfnamefont {J.}~\bibnamefont {{Wilson}}}, \bibinfo {author} {\bibfnamefont {K.~J.}\ \bibnamefont {{Millman}}}, \bibinfo {author} {\bibfnamefont {N.}~\bibnamefont {{Mayorov}}},
  \bibinfo {author} {\bibfnamefont {A.~R.~J.}\ \bibnamefont {{Nelson}}}, \bibinfo {author} {\bibfnamefont {E.}~\bibnamefont {{Jones}}}, \bibinfo {author} {\bibfnamefont {R.}~\bibnamefont {{Kern}}}, \bibinfo {author} {\bibfnamefont {E.}~\bibnamefont {{Larson}}}, \bibinfo {author} {\bibfnamefont {C.~J.}\ \bibnamefont {{Carey}}}, \bibinfo {author} {\bibfnamefont {{\.I}.}~\bibnamefont {{Polat}}}, \bibinfo {author} {\bibfnamefont {Y.}~\bibnamefont {{Feng}}}, \bibinfo {author} {\bibfnamefont {E.~W.}\ \bibnamefont {{Moore}}}, \bibinfo {author} {\bibfnamefont {J.}~\bibnamefont {{VanderPlas}}}, \bibinfo {author} {\bibfnamefont {D.}~\bibnamefont {{Laxalde}}}, \bibinfo {author} {\bibfnamefont {J.}~\bibnamefont {{Perktold}}}, \bibinfo {author} {\bibfnamefont {R.}~\bibnamefont {{Cimrman}}}, \bibinfo {author} {\bibfnamefont {I.}~\bibnamefont {{Henriksen}}}, \bibinfo {author} {\bibfnamefont {E.~A.}\ \bibnamefont {{Quintero}}}, \bibinfo {author} {\bibfnamefont {C.~R.}\ \bibnamefont {{Harris}}}, \bibinfo {author}
  {\bibfnamefont {A.~M.}\ \bibnamefont {{Archibald}}}, \bibinfo {author} {\bibfnamefont {A.~H.}\ \bibnamefont {{Ribeiro}}}, \bibinfo {author} {\bibfnamefont {F.}~\bibnamefont {{Pedregosa}}}, \bibinfo {author} {\bibfnamefont {P.}~\bibnamefont {{van Mulbregt}}},\ and\ \bibinfo {author} {\bibnamefont {{SciPy 1. 0 Contributors}}},\ }\bibfield  {title} {\bibinfo {title} {{SciPy 1.0: fundamental algorithms for scientific computing in Python}},\ }\href {https://doi.org/10.1038/s41592-019-0686-2} {\bibfield  {journal} {\bibinfo  {journal} {Nature Medicine}\ }\textbf {\bibinfo {volume} {17}},\ \bibinfo {pages} {261} (\bibinfo {year} {2020})},\ \Eprint {https://arxiv.org/abs/1907.10121} {arXiv:1907.10121 [cs.MS]} \BibitemShut {NoStop}%
\bibitem [{\citenamefont {Lewis}(2019)}]{Lewis_2019_GetDist}%
  \BibitemOpen
  \bibfield  {author} {\bibinfo {author} {\bibfnamefont {A.}~\bibnamefont {Lewis}},\ }\bibfield  {title} {\bibinfo {title} {{GetDist: a Python package for analysing Monte Carlo samples}},\ }\href {https://getdist.readthedocs.io} {\  (\bibinfo {year} {2019})},\ \Eprint {https://arxiv.org/abs/1910.13970} {arXiv:1910.13970 [astro-ph.IM]} \BibitemShut {NoStop}%
%%CITATION = ARXIV:1910.13970;%%
\bibitem [{\citenamefont {{Hunter}}(2007)}]{Hunter_2007_Matplotlib}%
  \BibitemOpen
  \bibfield  {author} {\bibinfo {author} {\bibfnamefont {J.~D.}\ \bibnamefont {{Hunter}}},\ }\bibfield  {title} {\bibinfo {title} {{Matplotlib: A 2D Graphics Environment}},\ }\href {https://doi.org/10.1109/MCSE.2007.55} {\bibfield  {journal} {\bibinfo  {journal} {Computing in Science and Engineering}\ }\textbf {\bibinfo {volume} {9}},\ \bibinfo {pages} {90} (\bibinfo {year} {2007})}\BibitemShut {NoStop}%
\bibitem [{\citenamefont {Cowan}\ \emph {et~al.}(2011)\citenamefont {Cowan}, \citenamefont {Cranmer}, \citenamefont {Gross},\ and\ \citenamefont {Vitells}}]{Cowan_2010_Asymptotic}%
  \BibitemOpen
  \bibfield  {author} {\bibinfo {author} {\bibfnamefont {G.}~\bibnamefont {Cowan}}, \bibinfo {author} {\bibfnamefont {K.}~\bibnamefont {Cranmer}}, \bibinfo {author} {\bibfnamefont {E.}~\bibnamefont {Gross}},\ and\ \bibinfo {author} {\bibfnamefont {O.}~\bibnamefont {Vitells}},\ }\bibfield  {title} {\bibinfo {title} {{Asymptotic formulae for likelihood-based tests of new physics}},\ }\href {https://doi.org/10.1140/epjc/s10052-011-1554-0} {\bibfield  {journal} {\bibinfo  {journal} {Eur. Phys. J. C}\ }\textbf {\bibinfo {volume} {71}},\ \bibinfo {pages} {1554} (\bibinfo {year} {2011})},\ \bibinfo {note} {[Erratum: Eur.Phys.J.C 73, 2501 (2013)]},\ \Eprint {https://arxiv.org/abs/1007.1727} {arXiv:1007.1727 [physics.data-an]} \BibitemShut {NoStop}%
\bibitem [{\citenamefont {Salucci}\ \emph {et~al.}(2025)\citenamefont {Salucci}, \citenamefont {Reiersrud}, \citenamefont {Read}, \citenamefont {Kvellestad},\ and\ \citenamefont {De~Bin}}]{Salucci:2025vem}%
  \BibitemOpen
  \bibfield  {author} {\bibinfo {author} {\bibfnamefont {C.~B.}\ \bibnamefont {Salucci}}, \bibinfo {author} {\bibfnamefont {H.~B.}\ \bibnamefont {Reiersrud}}, \bibinfo {author} {\bibfnamefont {A.~L.}\ \bibnamefont {Read}}, \bibinfo {author} {\bibfnamefont {A.}~\bibnamefont {Kvellestad}},\ and\ \bibinfo {author} {\bibfnamefont {R.}~\bibnamefont {De~Bin}},\ }\bibfield  {title} {\bibinfo {title} {{The asymptotic distribution of the likelihood ratio test statistic in two-peak discovery experiments}},\ }\href@noop {} {\  (\bibinfo {year} {2025})},\ \Eprint {https://arxiv.org/abs/2512.19333} {arXiv:2512.19333 [hep-ex]} \BibitemShut {NoStop}%
\bibitem [{\citenamefont {Chernoff}(1954)}]{Chernoff_1954_Distribution}%
  \BibitemOpen
  \bibfield  {author} {\bibinfo {author} {\bibfnamefont {H.}~\bibnamefont {Chernoff}},\ }\bibfield  {title} {\bibinfo {title} {On the distribution of the likelihood ratio},\ }\href {https://doi.org/10.1214/aoms/1177728725} {\bibfield  {journal} {\bibinfo  {journal} {The Annals of Mathematical Statistics}\ }\textbf {\bibinfo {volume} {25}},\ \bibinfo {pages} {573} (\bibinfo {year} {1954})}\BibitemShut {NoStop}%
\bibitem [{\citenamefont {Naredo-Tuero}\ \emph {et~al.}(2024)\citenamefont {Naredo-Tuero}, \citenamefont {Escudero}, \citenamefont {Fern{\'a}ndez-Mart{\'\i}nez}, \citenamefont {Marcano},\ and\ \citenamefont {Poulin}}]{Naredo-Tuero_2024_Critical}%
  \BibitemOpen
  \bibfield  {author} {\bibinfo {author} {\bibfnamefont {D.}~\bibnamefont {Naredo-Tuero}}, \bibinfo {author} {\bibfnamefont {M.}~\bibnamefont {Escudero}}, \bibinfo {author} {\bibfnamefont {E.}~\bibnamefont {Fern{\'a}ndez-Mart{\'\i}nez}}, \bibinfo {author} {\bibfnamefont {X.}~\bibnamefont {Marcano}},\ and\ \bibinfo {author} {\bibfnamefont {V.}~\bibnamefont {Poulin}},\ }\bibfield  {title} {\bibinfo {title} {{Critical look at the cosmological neutrino mass bound}},\ }\href {https://doi.org/10.1103/PhysRevD.110.123537} {\bibfield  {journal} {\bibinfo  {journal} {Phys. Rev. D}\ }\textbf {\bibinfo {volume} {110}},\ \bibinfo {pages} {123537} (\bibinfo {year} {2024})},\ \Eprint {https://arxiv.org/abs/2407.13831} {arXiv:2407.13831 [astro-ph.CO]} \BibitemShut {NoStop}%
\bibitem [{\citenamefont {Chebat}\ \emph {et~al.}(2026)\citenamefont {Chebat} \emph {et~al.}}]{Chebat_2025_Cosmological}%
  \BibitemOpen
  \bibfield  {author} {\bibinfo {author} {\bibfnamefont {D.}~\bibnamefont {Chebat}} \emph {et~al.},\ }\bibfield  {title} {\bibinfo {title} {{Cosmological neutrino mass: a frequentist overview in light of DESI}},\ }\href {https://doi.org/10.1088/1475-7516/2026/01/041} {\bibfield  {journal} {\bibinfo  {journal} {JCAP}\ }\textbf {\bibinfo {volume} {01}},\ \bibinfo {pages} {041}},\ \Eprint {https://arxiv.org/abs/2507.12401} {arXiv:2507.12401 [astro-ph.CO]} \BibitemShut {NoStop}%
\bibitem [{\citenamefont {Self}\ and\ \citenamefont {Liang}(1987)}]{Self_1987_Asymptotic}%
  \BibitemOpen
  \bibfield  {author} {\bibinfo {author} {\bibfnamefont {S.~G.}\ \bibnamefont {Self}}\ and\ \bibinfo {author} {\bibfnamefont {K.-Y.}\ \bibnamefont {Liang}},\ }\bibfield  {title} {\bibinfo {title} {Asymptotic properties of maximum likelihood estimators and likelihood ratio tests under nonstandard conditions},\ }\href@noop {} {\bibfield  {journal} {\bibinfo  {journal} {Journal of the American Statistical Association}\ }\textbf {\bibinfo {volume} {82}},\ \bibinfo {pages} {398} (\bibinfo {year} {1987})}\BibitemShut {NoStop}%
\end{thebibliography}%

\end{document}